\documentclass[aps,prb,twocolumn,preprintnumbers,amsmath,amssymb]{revtex4}

\usepackage{graphicx} % Include figure files
\usepackage{color}
\usepackage{dcolumn}% Align table columns on decimal point
\usepackage{bm}% bold math
\usepackage{physics}        
\usepackage{amsmath}
\usepackage{amssymb}
\usepackage{color}
\usepackage[normalem]{ulem}%Don't use underline
\usepackage{braket}
\usepackage{textcomp}

\newcommand{\bla}[1]{\textcolor{black}{#1}}

\begin{document}
\title{Proposed optomechanical systems based on luminescence-induced optical forces}

\author{Hideki Arahari${}^{1}$, Sota Konishi${}^{2}$\bla{, Kodai Takaoka${}^{2}$}, Seiji Akita${}^{2}$ and Hajime Ishihara${}^{1,3}$}
\affiliation{${}^{1}$Department of Material Engineering Science, Graduate School of Engineering Science, Osaka University, 1-3 Machikaneyama, Toyonaka, Osaka 560-8531, Japan}
\affiliation{${}^{2}$Department of Physics and Electronics, Osaka Metropolitan University, 1-1 Naka-ku, Sakai, Osaka 599-8531, Japan}
\affiliation{${}^{3}$Ritsumeikan Semiconductor Application research center (RISA), Ritsumeikan University, Kusatsu, Shiga 525-8577, Japan}
\date{\today}

%%%%%%% Abstract %%%%%%%
% \input{Abstract}
\begin{abstract}
We propose an optomechanical system utilizing luminescence-induced optical forces (\bla{LIOFs}). Anisotropic dielectric structures enhance the recoil force from the luminescence. The optomechanical resonator consists of a composite film with a dielectric membrane, luminescent nanofilm, and a metallic substrate. The \bla{LIOF} causes a mechanical frequency shift in the oscillator known as the optical spring effect. These results link the quantum properties of luminescent nanomaterials with those of other quantum-mechanical systems with vastly different frequency regimes via induced vibrational modes.
\end{abstract}

\pacs{}
\maketitle

%%%%%%%%%%%%Introduction%%%%%%%%%%%%%
Optical forces, which are used to manipulate objects at the atomic to micrometer scales, have been applied in various fields. Various techniques, such as optical tweezers \citep{Ashkin1986-yg} and atomic cooling \citep{Chu1985-gx}, have been employed in molecular biology \citep{Xin2020-hw,Corsetti2021-oq}, photochemistry \citep{Ito2011-tm, Cheng2020-kd}, and optomechanics \bla{\citep{Kippenberg2008-xj,Marquardt2009-vq,Aspelmeyer2014-fw}}. Recent advances have focused on structurally designed light fields, such as vortices \citep{Zhang2018-fq, Tamura2019-sm, Zhu2021-lx, Stilgoe2022-ir, Tao2023-yd}, localized surface plasmons \citep{Kotsifaki2019-oe, Zhang2021-ez, Hoshina2017-sf, Tanaka2020-gu, Shoji2020-sw, Fujiwara2021-vt}, and on the manipulation of light absorption and scattering owing to electronic resonant effects \citep{Iida2002-pu, Agayan2002-it, Iida2003-om, Li2006-ap, Skelton-Spesyvtseva2015-kp,Fujiwara2021-ld, Ishihara2021-xc}. Additionally, the luminescence-induced optical force (\bla{LIOF}) arises from both stimulated and spontaneous emissions. Although the stimulated recoil force has been studied \bla{\citep{Mizrahi2010-wu, Webb2011-cx, Kudo2012-ah, Bian2017-hi, Chen2020-hf, Kudo2023-as, Ito2024-vo, horai2025proposed}}, discussions on \bla{LIOF} attributed to spontaneous emissions have been limited. The symmetrical cancellation of \bla{LIOF} can be overcome by employing anisotropic emitter designs or dielectric environments, and the symmetrical cancellation of \bla{LIOF} can be overcome, thus making it a significant contributor to material motion.

One prominent application of the optical force is optomechanics \citep{Gigan2006-ao, Kleckner2006-nc, Arcizet2006-fh, Thompson2008-as, Groblacher2009-tf, Norte2016-gw, Stockill2019-qt, Barzanjeh2022-yp}. \bla{Here,} incident light induces the mechanical motion in mirrors that form cavity structures, coupling the optical cavity resonance and mechanical oscillations. In published optomechanics studies, the scattering force drives the oscillatory motion of the mirrors. 
However, \bla{the coupling of the coherent mechanical oscillator with the quantum mechanical emitter through \bla{LIOF} remains nontrivial and has not yet been explored.}
If \bla{LIOF} contributes to mechanical motion, the functionalities of optomechanics could be greatly enhanced. For example, this mechanism could directly link the emission behaviors, based on the quantum properties of the emitters, to other quantum systems \bla{across} vastly different frequency regimes via induced vibrational modes\bla{---even without photodetectors. Moreover,} the luminescent properties of quantized electronic systems can be modified by controlling their mechanical modes \bla{\citep{Ohta2021-zy,Spinnler2024-ay}}. \bla{Additionally,} optomechanics \bla{could be implemented} without external light sources. \bla{A particularly intriguing possibility is the synchronization \bla{\citep{Pikovsky2002-wa, Madiot2021-mh, Defoort2022-kq}} of incoherent emitters through coupled composite mechanical oscillators.}

To increase the widespread use of \bla{LIOF} optomechanics, one challenge is clarifying how the luminescence of materials can drive mechanical motion.      
To address this issue, we theoretically demonstrated the feasibility of luminescence-driven optomechanics by constructing a model with basic but realistic material components. Assuming that the technique can be used to fabricate a bridged thin layer as a mechanical oscillator \citep{Inoue2017-pk}, we can consider a cavity structure in which a luminescent film and a metallic mirror are arranged in parallel, as shown in Fig. \ref{fig1}. We numerically demonstrate that \bla{LIOF} modulates the mechanical frequency of the film owing to the cavity enhancement of the luminescent light, as shown in Fig. \ref{fig1}.
Luminescence-driven optomechanics has been reported to be active optomechanics \citep{Yu2022-no, Ge2013-ss} and laser optomechanics \citep{Yang2015-zc, Czerniuk2014-vy}. However, these studies considered single-photon modes, and the luminescence did not cause mechanical motion in the emitters but drove mirrors placed around the emitters. In this study, we evaluated \bla{LIOF} on emitters using the incoherent luminescence described based on the emission spectrum. 

%%%%%%% Figure 1 %%%%%%%
\begin{figure*}[htbp] % CCCCCCCCCCCCCCCCCCCCCC
  \includegraphics[width=\linewidth]{fig1.png}
  \vspace{-1.0em}
  \caption{
    Schematic view of the \bla{optomechanical resonator and its cross-sectional diagram. The system consists of a luminescent film ($\text{FAPbBr}_{3}$, thickness $d$) on a SiN membrane ($w$=50 nm) with a metallic mirror (Al) substrate. The excitation light is incident vertically, with two cases: (i) resonant excitation and (ii) non-resonant excitation. The spacer supports the membrane, which consists of the SiN membrane and the luminescent film, enabling it to act as a mechanical oscillator.} The size of the SiN membrane window is \bla{$S_{\text{SiN}} = 1$ mm $\times$ $1$ mm}. The parameters of the luminescent film (\bla{$\text{FAPbBr}_{3}$}) include the transverse energy and the longitudinal-transverse (LT) splitting energy of the exciton which are $\hbar \omega_{\bla{b}} = \bla{2.24}$ eV and $\Delta_{\text{LT}\bla{,b}} = \bla{3.4}$ meV, respectively \bla{\citep{Jain2021-jo,Kirstein2024-kn}}. The background dielectric constant $\varepsilon_{\text{b}}$ was set to \bla{4.89 \citep{Jain2021-jo}}. The optical parameters of the SiN membrane and Al substrate were also referenced from \citep{Philipp1973-gb, Baak1982-mh, Rakic1995-ew}.
      \label{fig1}
      }
\end{figure*} % CCCCCCCCCCCCCCCCCCCCCC

%%%モデル説明 Fig.1%%%
In the numerical demonstration, we assumed a square\bla{-type} optomechanical resonator formed by a luminescent nanofilm and a metallic mirror substrate, as shown in Fig. \ref{fig1}. \bla{Among various light-emitting materials, organic--inorganic perovskites have emerged as promising candidates due to their high PL quantum efficiency. In this study, we focus on $\text{FAPbBr}_{3}$ \citep{Trinh2018-ej,Tamarat2019-tk, Rubino2020-nj, Jain2021-jo, Hu2022-zt,Kirstein2024-kn} as the luminescent material.} The cavity structure is constructed by fabricating a luminescent film on a silicon nitride (SiN) membrane and transferring it onto an Al mirror substrate. In this way, the SiN membrane helps simplify the fabrication and transfer processes of the luminescent film and acts as an oscillator with low-mechanical damping characteristics \citep{Kleckner2011-nz, Kemiktarak2012-tm}. 
% \bla{Furthermore, the spacer functions as an optical waveguide for excitation. Future advancements could enable excitation through the spacer, eliminating the contribution of optical force from the excitation light to the oscillator.}
% \bla{Furthermore, optical force from the excitation light to the oscillator can be eliminated by utilizing the spacer supporting the oscillator (as shown in Fig. \ref{fig1}(b)) as an optical waveguide or through carrier injection.}
\bla{In this setup, the luminescent film does not need to cover} the entire area of the SiN membrane because if the diameter of the film is much larger than the light wavelength, the cavity works. Herein, the photoluminescence (PL) electric field in the cavity is enhanced or suppressed depending on the distance $L$ between the film and mirror owing to the optical confinement effect. The \bla{LIOF} exerted on the composite film was evaluated by calculating the PL electric field surrounding the film. 

%%%%%%%%%%%%Theory%%%%%%%%%%%%%
%%%%%PL Theory%%%%%
\bla{These calculations} were performed with reference to the PL theory of excitons in solids \citep{Matsuda2016-za} and the optical force theory based on Maxwell's stress tensor \citep{Iida2002-pu} (See Appendices A-C for details). The Hamiltonian considers a coupled system of excitons and radiation fields according to
\begin{align}
    \hat{H} =& \sum_{\eta} \hbar\Omega_{\eta} \hat{a}^{\dagger}_{\eta} \hat{a}_{\eta}
    + \bla{\sum_{o}^{b,c}\sum_{\mu=1}^{N_{o}} \hbar \Omega_{\mu, o}^{\text{ex}} \hat{o}_\mu^\dagger \hat{o}_\mu}
    \nonumber \\
    &-\bla{\sum_{o}^{b,c}} \int{\text{d} z} \hat{P}_{\text{ex}\bla{,o}}(z) \hat{E}(z),
\end{align}
where $\hat{a}^{\dagger}_{\eta} (\hat{a}_{\eta})$ denotes the creation (annihilation) operator of the $\eta$th photon mode with the energy $\hbar\Omega_{\eta}$ \bla{of entire system including the oscillating film and substrate}, and \bla{$\hat{o}^{\dagger}_{\mu} (\hat{o}_{\mu})$} represents the creation (annihilation) operator of the $\mu$th exciton state. \bla{We assumed two modes for the excitons: emission levels $\hat{b}_{\mu}$ and excitation levels $\hat{c}_{\mu}$.} $\hat{P}_{\text{ex}\bla{,o}}(z)=\sum_{\mu} (P_{\mu\bla{,o}}(z)\bla{\hat{o}_{\mu}}(t) + \text{H.c.}$) is the excitonic polarization operator, and $\hat{E}(z)$ is the electric field. As the center-of-mass motion of excitons is confined to the film's thickness direction ($z$-direction), the eigenenergy for each exciton mode is expressed as $\hbar\Omega_{\mu\bla{,o}}^{\text{ex}} = \hbar\omega_{\bla{o}} + \hbar^2 K_{\mu}^2/(2m_{\text{ex}})$, where $K_{\mu} = \mu\pi/d$ is the quantized wavenumber with $\mu = 1, \cdots \bla{N_{o}}$. \bla{The excitons in the $\text{FAPbBr}_{3}$ film were almost bound to each site; therefore, we can safely approximate the translational mass of the exciton $m_{\text{ex}}$ along the film axis as $m_{\text{ex}} \rightarrow \infty$ (See Appendix A for details).} We derived quantum master equations for excitons by considering nonradiative decay and dephasing processes. The equations were solved using the quantum Maxwell’s equation $\hat{E}(z, t)=\hat{E}_{0}(z, t)+\sum_{\bla{o,}\mu} \int \mathrm{d} z^{\prime} G_{\bla{o}}\left(z, z^{\prime}\bla{, \omega_o}\right) P_{\mu\bla{,o}}\left(z^{\prime}\right) \bla{\hat{o}_{\mu}}(t)$, expressed using Green's function $G_{\bla{o}}\left(z, z^{\prime}\right)$ self-consistently \citep{Matsuda2016-za}, where the Green's function reflects the spatial structure of the background dielectric constant (including metals) \bla{determining the photon modes \{$\eta$\}} in an assumed system (Fig. \ref{fig1} in this study) \citep{Chew1995-hr}. By performing the Fourier transform under steady-state conditions, we calculated the \bla{resonance} PL spectrum: $S_{\mathrm{inc}}(z,\omega) = 1/\pi \sum_{\mu, \mu^{\prime}} \operatorname{Re}\left[\int_{0}^{\infty} \mathrm{d} \tau \braket{\Delta \hat{E}^{\dagger}(z,0)\Delta\hat{E}(z,\tau)} e^{i \omega \tau}\right]$, where we treated the incoherent component of the electric field intensity $\braket{\Delta\hat{E}^{\dagger}(z,0)\Delta\hat{E}(z,\tau)} = \braket{\hat{E}^{\dagger}(z,0)E(z,\tau)} - \braket{\hat{E}^{\dagger}(z,0)}\braket{\hat{E}(z,\tau)}$ as the PL intensity \citep{Carmichael2002-aq}. \bla{In the case of non-resonant excitation, the PL spectrum is expressed as $S_{\mathrm{inc}}(z,\omega) = 1/\pi \sum_{\mu, \mu^{\prime}} \operatorname{Re}\left[F_{\mu}^*(z)F_{\mu'}(z)\int_{0}^{\infty} \mathrm{d} \tau \braket{\hat{b}_{\mu}^{\dagger}(0)\hat{b}_{\mu'}(\tau)} e^{i \omega \tau}\right]$, where $F_{\mu}(z) = \int{\mathrm{d}z'}G(z,z',\omega)P_{\mu,b}(z')$. (See Appendices A and B for details on excitonic parameters and the number of exciton and photon modes considered.)}

%%%%%OF Theory%%%%%
The time-averaged optical force $\braket{f_z}$ exerted on the film can be expressed as follows. Considering only the force acting perpendicular to the surface of the film (in the $z$ direction), $\expval{f_z(\omega)} = \frac{S}{2}\varepsilon_0 ( \abs{E_\text{L1}(\omega)}^2 + \abs{E_\text{L2}(\omega)}^2 - \abs{E_\text{U1}(\omega)}^2 - \abs{E_\text{U2}(\omega)}^2)$, 
where $E_{\text{L1}}(E_{\text{L2}})$ represents the electric field with an upward (downward) wavenumber on the lower surface of the film, as shown in Fig. \ref{fig1}, and $E_{\text{U1}}(E_{\text{U2}})$ represents that with an upward (downward) wavenumber at the upper surface of the film. $S$ is the area of the film and $\varepsilon_0$ is the vacuum permittivity. 
To calculate \bla{LIOF} $\braket{F_z^{\text{inc}}}$, we used the PL intensity obtained by the previously described process.
$\abs{E_\text{U1}(\omega)}^2$ is the PL intensity $S_{\mathrm{inc}}(\bla{z_{\text{up}}},\omega)$ at the upper surface of the film and $\abs{E_\text{U2}(\omega)}^2$ is \bla{zero. (For obtaining} $\abs{E_\text{L1}}^2 + \abs{E_\text{L2}}^2$, see Appendix C). Finally, \bla{LIOF} was obtained by integrating $\braket{f_{z}^{\text{inc}}(\omega)}$ over the emission frequency $\braket{F_z^{\text{inc}}} = \int \braket{f_{z}^{\text{inc}}(\omega)} \text{d}\omega$. 

%%%%%%%%%%%%Results and Discussion%%%%%%%%%%%%%
%%%%%%% Figure 2 %%%%%%%
\begin{figure}[htbp] % CCCCCCCCCCCCCCCCCCCCCC
    \centering
    \includegraphics[width=\linewidth]{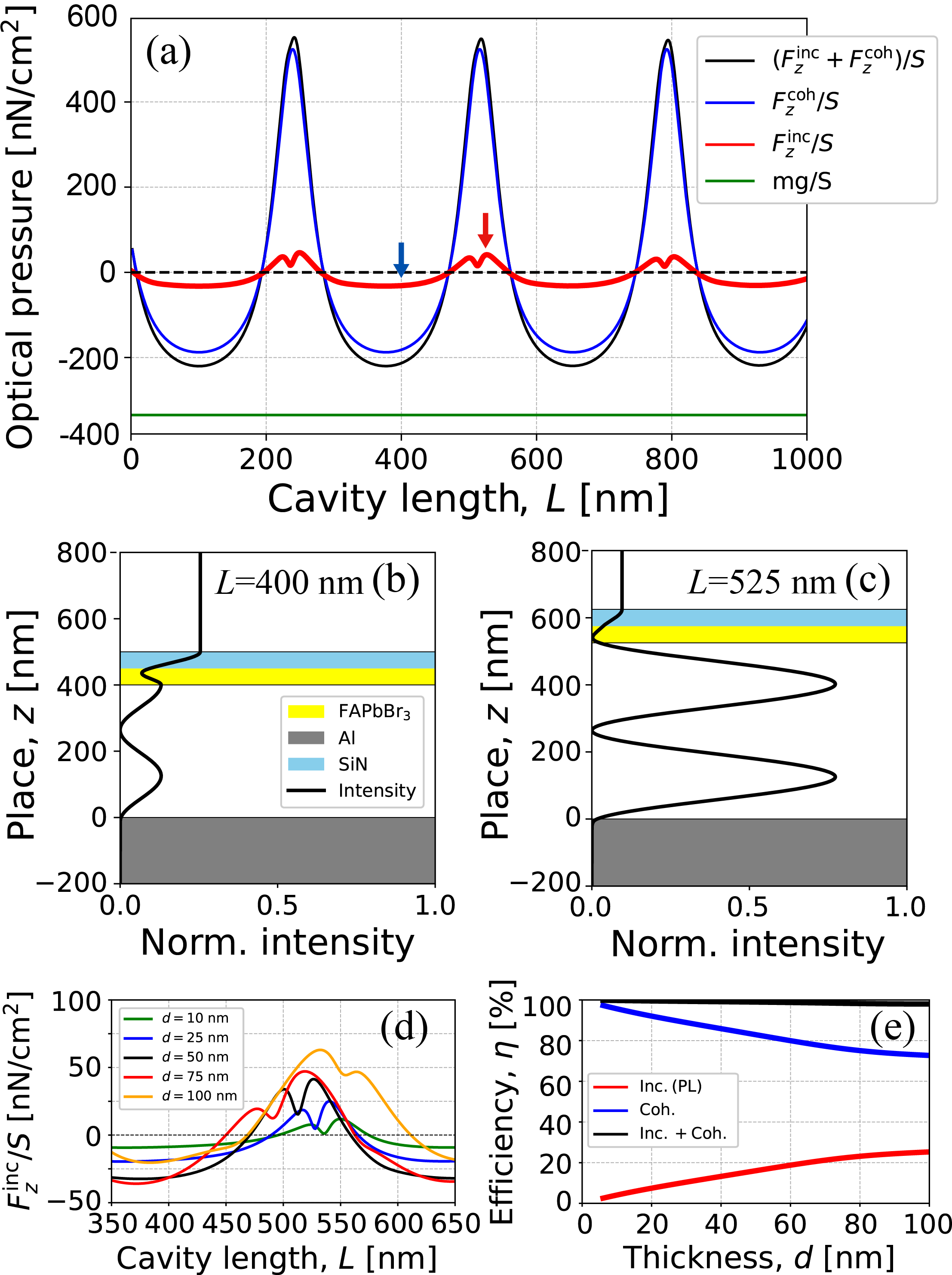}
    \caption{      
    (a) \bla{Optical force as a function of the cavity length $L$, with excitation light incident vertically onto the membrane at an intensity of $I = 50$ $\text{W}/\text{cm}^2$} (the meanings of line colors are explained in the main text). The results are for the luminescent film with $d = 50$ nm. 
    (b, c) Spatial distribution of the \bla{integrated} photoluminescence (PL) intensity at $L = \bla{400}$ nm (\bla{525} nm)\bla{, corresponding to conditions} where the \bla{LIOF} acts as an attractive (repulsive) force, represented by a blue (red) arrow in Fig. \ref{fig2}(a). The blue (yellow, gray) region represents the SiN (\bla{$\text{FAPbBr}_{3}$}, Al).
    (d) $L$-dependence of \bla{LIOF for} various film thicknesses. (e) \bla{Thickness} dependence of PL (incoherent) and coherent light efficiency \bla{in the absence of a mirror}. Herein, the nonradiative decay constant \bla{is $\hbar\gamma_{b}=0.2$ {\textmu}eV \citep{He2018-nd} and the dephasing constant $\hbar \Gamma_{b}=50$ meV, which match the experimental absorption peak \citep{Trinh2018-ej} and the FWHM ($\sim 12$ nm, \citep{Hu2022-zt}) of PL spectrum at cryogenic condition (See Appendix A for details).}
    \label{fig2}
        }
\end{figure} % CCCCCCCCCCCCCCCCCCCCCC

%Fig2説明
First, we assumed a case in which the resonant excitation light was irradiated vertically from the top part of the SiN membrane in Fig. \ref{fig1}\bla{(i)}. The optical force owing to the coherent components of light, including the excitation light, can be evaluated as $\braket{F_z^{\text{coh}}} = \braket{f_{z}^{\text{coh}}(\omega_{\text{in}})}$.
Fig. \ref{fig2}(a) shows the cavity length $L$ dependence of the optical force exerted on the composite film (luminescent film with a thickness $d = 50$ nm and a SiN membrane). To excite the luminescent film, we assumed steady irradiation of electronic resonant light ($\hbar\omega_{\text{in}}=\bla{\hbar\omega_{b}=2.24}$ eV) with an intensity of $I = \bla{50}$ $\text{W/cm}^2$\bla{, which is near the limit of the linear response region \citep{Trinh2018-ej},} from the top of the film. The red and blue lines in Fig. \ref{fig2}(a) represent the \bla{LIOF} and the optical force induced by the coherent component of light, including the excitation light, respectively, and the black line represents their sum.
\bla{The optical confinement effect of the resonator causes periodic enhancement of \bla{LIOF}, repeating every half-wavelength of the emission ($\lambda_{\text{PL}}/2 \sim$ 277 nm)}. This is due to the single-peak-like PL spectrum of \bla{$\text{FAPbBr}_{3}$} (See Appendices A and B). \bla{Under resonant excitation, the optical force induced by coherent light also exhibits similar periodic enhancement.}
Figures \ref{fig2} (b, c) show the spatial distributions of the integrated PL intensity at the cavity length where (b) \bla{suppression} and (c) \bla{enhancement} of \bla{LIOF} occurs. The \bla{latter} shows strong confinement of photons in the cavity. 
The black line in Fig. \ref{fig2}(a) shows the sum of the \bla{LIOF} and optical force induced by the coherent light. \bla{The condition} $L \sim \bla{281 + n\lambda_{\text{PL}}/2}$ nm, \bla{where $n$ is a non-negative integer, represents} stable equilibrium positions near which the optical force acts as a restoring force (optical trapping of the film).  
The enhanced \bla{optical force} is greater than the gravitational force of the composite film per unit area (green line in Fig. \ref{fig2}(a), $mg/S \simeq \bla{335}$  $\text{\bla{n}N}/\text{cm}^2$). This result indicates that the film could be levitated by \bla{optical force, leading} to ideal optomechanical systems with high mechanical quality factors \bla{($Q_{\text{m}}$)} due to no clamping loss, such as optically trapped mirrors \citep{Singh2010-ny} and nanoparticles \citep{Gieseler2013-dm, Delic2020-tj}. % Considering only the optical force induced by coherent light, as indicated by the blue line in Fig. \ref{fig2}(a), no stable equilibrium position was found. 
% \bla{At present, the optical force exerted by coherent light dominates over LIOF. However, in the future, the optical force from the excitation light could be eliminated by utilizing the spacer supporting the oscillator (as shown in Fig. \ref{fig1}(b)) as an optical waveguide or by employing carrier injection.}

%Fig2(d,e)説明
Figure \ref{fig2}(d) shows the \bla{LIOF} with thicknesses ranging from $d=10$ to $100$ nm. \bla{The} results show that we can obtain enhanced \bla{LIOF} on any film thickness in this range. \bla{Within this range, \bla{the external} quantum efficiency (\bla{EQE}) increases with increasing luminescent film thickness, leading to a corresponding enhancement of the optical force.}
% This is because the PL quantum efficiency (PLQE) remains relatively unchanged in this range, as shown in Fig. \ref{fig2}(e). 
The \bla{EQE} is evaluated as $\eta = \int S_{\text{inc}}(\omega) \mathrm{d}\omega/\abs{E_0}^2$, where $\abs{E_0}^2 = 2I/(c\varepsilon_0)$ and $c$ is the speed of light. The efficiency of the coherent-light component can be expressed as $\abs{E(\omega_{\text{in}})}^2/\abs{E_0}^2$. A structural dip is formed in the enhanced \bla{LIOF} when the film thickness is small because the emitter is placed at the nodes of the standing wave of the excitation light, and the emitter is not sufficiently excited \bla{and consequently does not emit light. (See Appendix E.)}

\begin{figure}[htbp] % CCCCCCCCCCCCCCCCCCCCCC
        \centering
          \includegraphics[width=\linewidth]{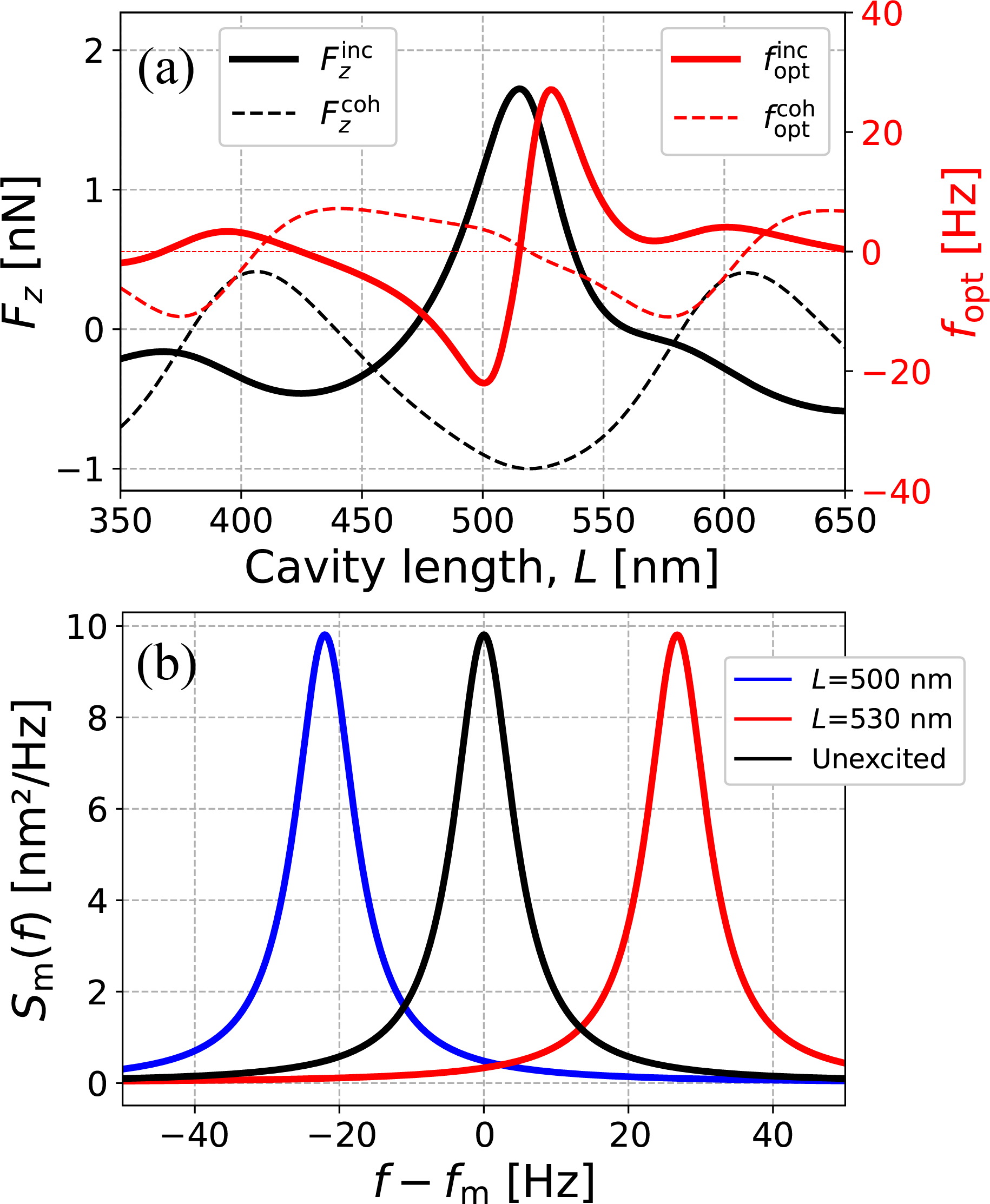}
    \caption{
    (a) (Black line) $L$-dependence of \bla{optical forces} and (Red line) Mechanical frequency shift induced by the optical spring effect \bla{under non-resonant excitation, as shown in Fig. \ref{fig1}(ii). The solid (dotted) line represents LIOF (optical force by coherent light). The experimental setup assumes a square-type optomechanical resonator with a luminescent film of thickness $d = 50$ nm irradiated by excitation light ($\hbar\omega_{\text{in}}=\hbar\omega_{c}=3.06$ eV) at an intensity of $I = 50$ $\text{W/cm}^2$.}
    (b) Power spectral density (PSD) of the LIOF at cavity lengths $L = 500$ nm (blue), $530$ nm (red). The black line shows the case where no emission occurs due to no excitation of the luminescent film. \bla{Herein, the damping constants of the radiative levels are consistent with those in Fig. \ref{fig2}.
    As for the parameters of the excited states, the dephasing $\hbar \Gamma_c=0.4$ meV \citep{Tamarat2019-tk} at cryogenic condition. The oscillator strength is set as $\Delta_{\text{LT},c} = 0.03\Delta_{\text{LT},b}$ based on the experimental absorption ratio \citep{Trinh2018-ej}. (See Appendix A for details.) $\left|F_{z, 0}^{\text{act}}\right|$ is set to 40 pN in our calculation.}
    }         \label{fig3}
\end{figure}

%%%%%%% Figure 3 %%%%%%%
% In the previous test, the film was subjected to an optical force by the excitation light in addition to the \bla{LIOF}. However, in the presence of the \bla{LIOF}, mechanical systems operate without direct light irradiation. Thus, in Fig. \ref{fig3} and thereafter, we considered the situation where the excitation light does not exert an optical force by providing a steady excitation energy to the luminescent films. For operations corresponding to these situations, we can consider excitations using an optical waveguide (forming the spacer supporting the oscillator), as shown in Fig. \ref{fig1}(b), or carrier injection that will be realized by techniques which will be introduced in the future. For simplicity, in this study, the exciton--polarization interaction energy, $\int{P_{\mu}^*(z)E_0(z, I)}\text{d}z$, has a constant value corresponding to the excitation light intensity $I$ in the calculation. 
In the previous test, \bla{the optical force during resonance PL was evaluated. In this case, since the wavelengths of the PL and excitation light are nearly identical, the enhancement of the optical force occurs under the same cavity conditions as depicted in Fig. \ref{fig2}(a). Furthermore, under these conditions, the membrane is positioned near the nodes of the standing wave of the excitation light, reducing excitation and optical absorption. As a result, the LIOF enhancement is smaller compared to the optical force from coherent light, making it challenging to distinguish the LIOF.
To address this, from Fig. \ref{fig3} onwards, the excitation states $\ket{c_{\mu}}$ of the luminescent film are assumed to be populated and subsequently relax to the radiative levels $\ket{b_{\mu}}$ at each radiative rate, incorporating the Stokes shift for the calculation of luminescence (non-resonant PL). This approach enables the clear observation of the LIOF since the resonance conditions in the cavity differ between the excitation and emission wavelengths.}
In Fig. \ref{fig3}(a), the black line shows the cavity length $L$ dependence of \bla{optical forces under non-resonant excitation ($\lambda_{\text{in}} = 405$ nm).} \bla{The \bla{LIOF} is enhanced at a specific cavity length, as discussed in Fig. \ref{fig2}(a),} and this \bla{LIOF} enhancement occurs at \bla{a period of $\lambda_{\text{PL}}/2$}. \bla{It can be seen that the cavity conditions for the enhancement of \bla{LIOF} and optical force induced by the excitation light are misaligned.}
Herein, we evaluated the contribution of \bla{optical force} to the oscillator motion using the optical spring effect \citep{Sheard2004-tv, Duy-Vy2013-th}, which is the mechanical frequency shift of the oscillator due to \bla{optical force} that varies with the cavity length of the resonator. The optical spring constant $k_{\text{opt}}$ is defined as,
$k_{\text{opt}}(L) \equiv - \mathrm{d} \ev{F_{z}(L)}/\mathrm{d}L$ \citep{Sheard2004-tv}.
We can evaluate the optomechanically modified effective resonance frequency $f_{\text{eff}}(=\omega_{\text{eff}}/2\pi)$ and damping $\Gamma_{\text{eff}}$ as \citep{Metzger2004-nx,Metzger2008-ww,Usami2012-vi}
\begin{align}
   f_{\text{eff}} &= f_{\text{m}} \sqrt{ 1 + \frac{1}{1+(\omega\tau)^2}\frac{k_{\text{opt}}}{m_{\text{eff}}\omega^2_{\text{m}}} }, \label{eq:5} \\
   \Gamma_{\text{eff}} &= \Gamma_{\text{m}}\left( 1 - Q_{\mathrm{m}} \frac{\omega_{\mathrm{m}} \tau}{1+(\omega \tau)^2} \frac{k_{\text{opt}}}{m_{\text{eff}}\omega^2_{\text{m}}} \right). \label{eq:6}
\end{align}
where $m_{\text{eff}}$ is the effective mass of the oscillator and $\tau$ is the delay time of the optical force response. $f_{\text{m}}(=\omega_{\text{m}}/2\pi)$ is the mechanical frequency of the resonator and $\Gamma_{\text{m}}$ is the mechanical damping constant, which is obtained as $\Gamma_{\text{m}}=\omega_{\text{m}}/Q_{\text{m}}$. In the present model, we can treat $\tau \sim 0$ because the Q-value of the optical cavity is small. The mechanical frequency shift $f_{\text{opt}}$ due to the optical spring effect is $f_{\text{opt}}(L) = f_{\text{eff}} - f_{\text{m}}$. In Fig. \ref{fig3}(a), the red line shows the result obtained by re-expressing the black line considering the $L$-dependence of the mechanical frequency shift due to the optical spring effect. The shift is negative (positive) in the region where the cavity length is smaller (larger) than that at the peak of optical-force enhancement. 

%%%%%%% Figure 3 %%%%%%%
 % CCCCCCCCCCCCCCCCCCCCCC

%Fig3(b)説明
In Fig. \ref{fig3}(b), we \bla{treated the square-type mechanical oscillator as a one-dimensional damped harmonic oscillator and calculated the power spectral density (PSD) of the oscillator} using the \bla{LIOF} shown in Fig. \ref{fig3}(a). Here, we assumed that the mechanical oscillation was amplified \bla{by external actuation. Possible actuation methods include, for example, the irradiation of an actuation laser \citep{Inoue2017-pk} onto the Si frame surrounding the SiN membrane or electrical actuation \citep{Chowdhury2017-zn,Yoshikawa2019-pp}.} The PSD $S_{\text{m}}(\omega)$ is expressed as (see Appendix D),
\begin{align}
    S_{\mathrm{m}}(\omega) = \frac{1}{m_{\mathrm{eff}}^2} \frac{\left|F_{z, 0}^{\text{act}}\right|^2}{\left(\omega_{\mathrm{eff}}^2-\omega^2\right)^2+\left(\omega \Gamma_{\mathrm{eff}}\right)^2}, 
\end{align}
where $F_{z, 0}^{\text{act}}$ denotes the amplitude of the actuation force. The densities of the luminescent film and $\text{Si}\text{N}$ membrane are $\rho_{\bla{\text{FAPbBr}_{3}}}=3.67$ $\text{g}/\text{cm}^3$ \citep{Giri2020-qk} and $\rho_{\text{SiN}}=2.80$ $\text{g}/\text{cm}^3$ \citep{Serra2018-kw} respectively, the effective mass of the oscillator was set to $m_{\text{eff}}=(\rho_{\bla{\text{FAPbBr}_{3}}}S_{\bla{\text{FAPbBr}_{3}}}d + \rho_{\text{SiN}}S_{\text{SiN}}w)/4$, where $S_{\bla{\text{FAPbBr}_{3}}}$($S_{\text{SiN}}$) is the area of the $\bla{\text{FAPbBr}_{3}}$ \bla{(SiN)} film.
In this calculation, we assumed $f_{\text{m}} = \bla{400}$ \bla{k}Hz and $Q_{\text{m}} = \bla{3 \times 10^4}$ (these parameters were estimated using an actual experimental setup).
The mechanical frequency of the oscillator shifted negatively ($L = \bla{500}$ nm) and positively ($L = \bla{530}$ nm) from the given parameter $f_{\text{m}}$, thus indicating the optical spring effect shown in Fig. \ref{fig3}(a). This result demonstrates that the \bla{LIOF} effect on vibrational motion can be observed by detecting the optical spring effect in the experiments.
\bla{In the future, the optical force induced by the excitation light could be eliminated by utilizing the spacer supporting the oscillator (as shown in Fig. \ref{fig1}(b)) as an optical waveguide or by employing carrier injection.}
\bla{During excitation,} photothermally induced forces cause mechanical frequency shifts. However, the optical spring effect can be distinguished easily because the photothermal effect appears with a significantly longer delay time than the response time of the optical force \citep{Metzger2008-ww}. Experimental studies have also reported the separate detection of optical force and photothermal contributions \citep{Yamanishi2021-ya}.

%%%%%%%%%%%%Conclusion%%%%%%%%%%%%%
\bla{The results of this study} open up a new research field for luminescence-driven optical manipulation (based on the design of the dielectric environment of targeted systems) in contrast to the conventional approach that has been based on the spatial design of light fields. In addition, this research leads to mechanisms that convert luminescence modes---that reflect the material quantum properties---into mechanical modes. Recent studies demonstrated the control of quantum states within materials (e.g., electronic systems \bla{\citep{Ohta2021-zy,Spinnler2024-ay}} and exciton-polaritons \bla{\citep{Vyatkin2021-xz, Carlon-Zambon2022-mp, Shishkov2024-hc, Shishkov2024-ll}}) based on couplings with optomechanical systems. In contrast, our proposed luminescence-driven optomechanical system enables mutual control between the emitters and mechanical modes, potentially achieving self-narrowing of the emission spectra and self-amplification of the luminescence. Coupling the system with other quantum systems, such as magnons \citep{Zhang2016-dp, Li2018-el, Zhu2020-aw, Qi2021-gs, Shen2022-rg}, may allow access to the emitter quantum properties without photodetection, thus benefiting quantum property processing and quantum transducers. Although our demonstration focused on a specific material, similar systems can be realized using other luminescent materials (e.g., quantum dots and molecular materials) with different geometries. Given the relatively unexplored nature of phenomena involving \bla{LIOF}, these research topics have considerable importance.

%%%%%%Acknowledgement%%%%%
We thank T. Matsuda for the fruitful discussions. This study was supported in part by JSPS KAKENHI (Grant No. JP16H06504) for Scientific Research on Innovative Areas “Nano-Material Optical-Manipulation,” by JSPS KAKENHI (Grant Nos. JP21H05019 and JP21K18193), JST, and the establishment of University fellowships for the creation of science technology innovation (Grant No. JPMJFS2125).

\onecolumngrid

\appendix
\section{Resonance PL Theory for luminescent film}\label{Sec:Res_PL}
In this study, we employ the exciton photoluminescence (PL) theory in solids \citep{Matsuda2016-za} and optical force theory based on Maxwell's stress tensor \citep{Iida2002-pu}. This section describes the theoretical framework used to calculate the linear optical response of excitons. The Hamiltonian considers a coupled system of the radiation field and excitons whose center-of-mass motions are confined in the thickness direction ($z$-direction) in a thin film as follows,
\begin{align}
    \hat{H} = \sum_{\eta} \hbar\Omega_{\eta} \hat{a}^{\dagger}_{\eta} \hat{a}_{\eta} + \sum_{\mu} \hbar\Omega_{\mu}^{\text{ex}} \hat{b}^{\dagger}_{\mu} \hat{b}_{\mu} -\int{\text{d} z} \hat{P}_{\text{ex}}(z) \hat{E}(z), \label{Seq:1}
\end{align}
where \bla{$\hat{a}^{\dagger}_{\eta} (\hat{a}_{\eta})$ represents the creation (annihilation) operator of the $\eta$th photon mode with the energy $\hbar\Omega_{\eta}$ of entire system including the oscillating film and substrate, and $\hat{b}^{\dagger}_{\mu} (\hat{b}_{\mu})$ denotes the creation (annihilation) operator of the $\mu$th exciton state. $\hat{E}(z) \left( =\sum_{\eta} \alpha_\eta \lbrace i \mathcal{E}_{\eta}(z) \hat{a}_{\eta} + \mathrm{H.c.} \rbrace , \alpha_\eta=\sqrt{\hbar\omega_{\eta}/2\varepsilon_0} \right) $ is the electric field operator and $\mathcal{E}_{\eta}(z)$ is an eigenfunction of Maxwell equation, $\nabla \times \nabla \times \mathcal{E}_{\eta}(z)-\varepsilon_{\mathrm{b}}(z) \frac{\omega_\eta^2}{c^2} \mathcal{E}_{\eta}(z)=0$.}
$\hat{P}_{\text{ex}} \left(=\sum_{\mu} P_{\mu}(z)\hat{b}_{\mu} + \mathrm{H.c.} \right)$ is the polarization operator. Let $P_{\mu}(z)$ be expressed as $P_{\mu}(z)=P\psi_{\mu}(z)$ using $P^2 = \varepsilon_0\varepsilon_{\mathrm{b}}\Delta_{\text{LT},b}$ and let the wave function of excitonic center-of-mass motion $\psi_{\mu}(z)$ be expressed as $\psi_{\mu}(z) = \sqrt{2/d}\sin{K_{\mu}z}$ in quantized form with the $\sin$ function ($K_{\mu} = \mu\pi/d, \mu = 1,2,\cdots, \bla{N_b}$). \bla{Here, $\mu$ represents the order of exciton modes. In our calculations, we considered up to the 10th mode ($N_b=10$), which is deemed sufficient given that the thickness of the luminescent film is small compared to the wavelength of light (See Figure \ref{L-OF_n=9or11} for details). We confirmed that the results remained nearly unchanged even when the maximum order was varied. }
The exciton eigenenergy $\hbar\Omega_{\mu}^{\text{ex}}$ can also be expressed using the transverse-wave exciton energy $\hbar\omega_{\bla{b}}$ as $\hbar\Omega_{\mu}^{\text{ex}}=\hbar\omega_{\bla{b}}+\hbar^2 K_{\mu}^2/(2m_{\text{ex}})$. \bla{In semiconductor materials containing organics, excitons are typically bound to each site; thus, we treated the center-of-mass of the exciton $m_{\text{ex}}$ as effectively infinity. For example in PEPI, $m_{\text{ex}}$ is approximately $10^3 m_0$ \citep{Zhang2022-ve}, with $m_0$ being the electron mass, indicating that the mass is so large that excitons hardly move within the film. (In our previous studies on the luminescence of CuCl thin films \citep{Matsuda2016-za}, the exciton center-of-mass was finite, where $m_{\text{ex}} =2.3 m_0$).}

\bla{In this study, we operate within the weak coupling regime due to the relatively thin film thickness. In this regime, the coupling strength can be evaluated using Fermi's golden rule, which differs from the polariton regime [3] in which multiple energy exchanges occur between the photon and emitter. It is important to note that although our model includes a thin film and a metallic substrate forming a cavity, the film itself is not positioned within the cavity, thus it does not strongly interact with the cavity photons.}

\begin{figure}[htbp] % CCCCCCCCCCCCCCCCCCCCCC
        \centering
          \includegraphics[width=90mm]{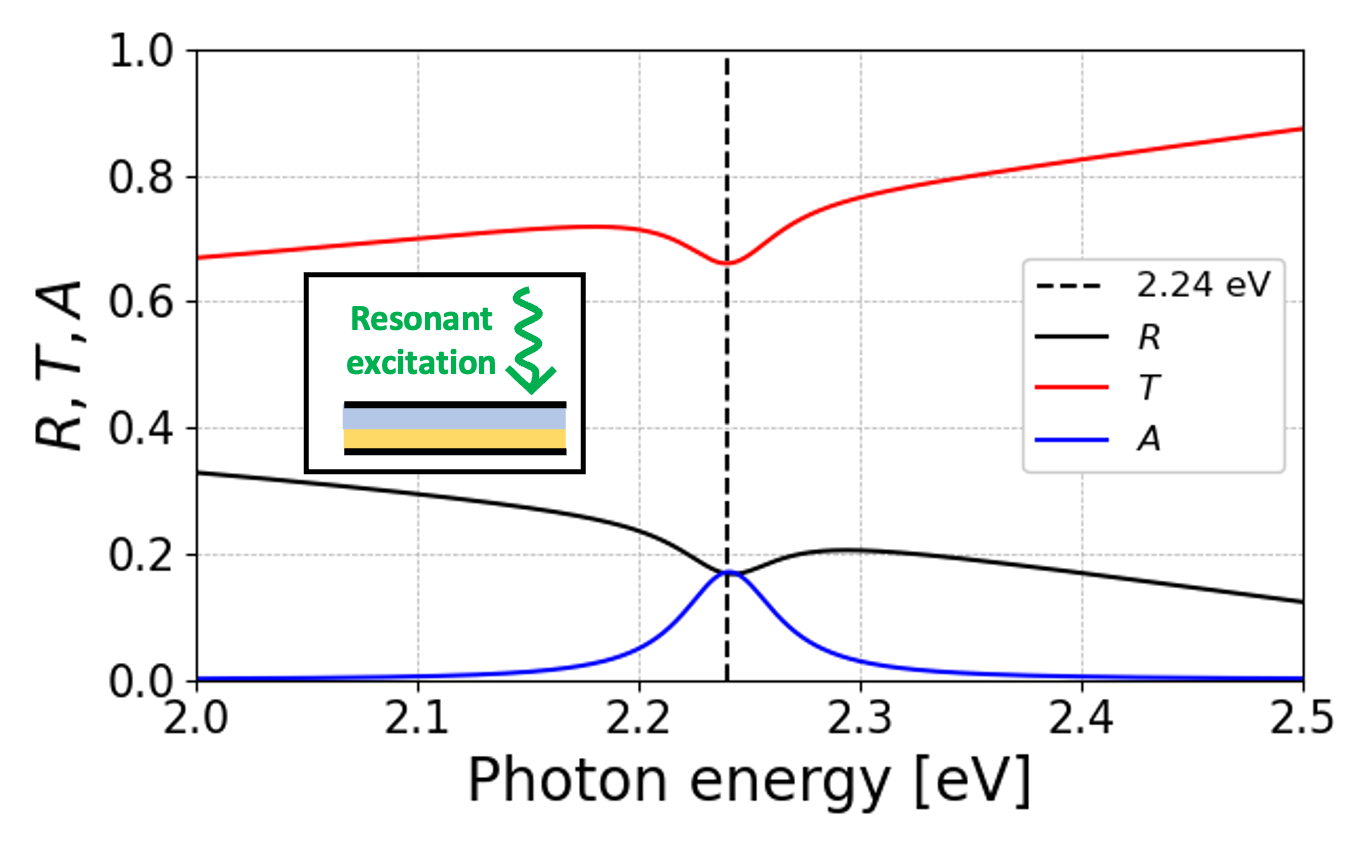}
    \caption{
    Reflection (black line: R), transmission (red line: T), and absorption (blue line: A) spectra of the composite film of SiN \bla{and $\text{FAPbBr}_{3}$ layer with thickness $w = d = 50$ nm under resonant excitation}. The black dotted line represents the excitation light energy $\hbar\omega_{\text{in}}$ (which was set in this paper to the transverse-wave exciton energy $\hbar\omega_{\text{in}}=\bla{\hbar\omega_{b}=2.24}$ \bla{eV} \citep{Jain2021-jo}). \bla{Herein, the LT splitting energy of excitons $\Delta_{\text{LT},b} = 3.4$ meV \citep{Kirstein2024-kn} and dephasing constant $\hbar \Gamma_{\text{ex}}=50$ meV are used in this paper.}
        \label{Optical_spe_res}
        }
\end{figure}

Nonradiative relaxation processes of excitons are treated by the quantum master equation based on the Born--Markov approximation \citep{Carmichael2002-aq} as,
\begin{align}
	&\frac{\partial \hat{\rho}(t)}{\partial t}=\frac{1}{i\hbar}[\hat{H},\hat{\rho}(t)] + \hat{L}_{\text{damp}}\hat{\rho}(t) + \hat{L}_{\text{phase}}\hat{\rho}(t), \label{Seq:2}\\
        &\hat{L}_{\text{damp}}\hat{\rho}(t) = \sum_{\mu}\frac{\gamma_{\text{ex}}}{2}[2\hat{b}_{\mu}\hat{\rho}(t)\hat{b}^{\dagger}_{\mu} - \{\hat{b}^{\dagger}_{\mu}\hat{b}_{\mu},\hat{\rho}(t)\} ], \label{Seq:3}\\
        &\hat{L}_{\text{phase}}\hat{\rho}(t) = \sum_{\mu}\frac{\Gamma_{\text{ex}}}{2}\big[[\hat{b}^{\dagger}_{\mu}\hat{b}_{\mu},\hat{\rho}(t)],\hat{b}^{\dagger}_{\mu}\hat{b}_{\mu} \big], \label{Seq:4}
\end{align}
where \bla{$\gamma_{\text{ex}}$ and $\Gamma_{\text{ex}}$ are the nonradiative decay and dephasing rate, respectively. In this paper, $\hbar\gamma_{\text{ex}} = 0.2$ $\mu$eV \citep{He2018-nd}.} From Equation (\ref{Seq:2}), the Heisenberg equation for the expected value of each exciton operator can be obtained \bla{as follows.
\begin{align}
    \frac{\mathrm{d}\braket{\hat{a}_\eta(t)}}{\mathrm{d} t} 
    &= -i \Omega_\eta \braket{\hat{a}_\eta(t)} + \frac{\alpha_\eta}{\hbar} \sum_\mu \int \mathrm{d}z \mathcal{E}_\eta^*(z) P_\mu(z) \braket{\hat{b}_\mu(t)}, \\
    \frac{\mathrm{d}\braket{\hat{b}_\mu(t)}}{\mathrm{d} t}  
    &= -(i\Omega_{\mu}^{\text{ex}}+\frac{\gamma_{\text{ex}}}{2}+\frac{\Gamma_{\text{ex}}}{2}) \braket{\hat{b}_\mu(t)} - \frac{1}{\hbar} \int \mathrm{d}z \sum_\eta \alpha_\eta P_\mu^*(z) \mathcal{E}_\eta(z) \braket{\hat{a}_\eta(t)}.
\end{align}
Herein, $\eta$ considers all the photon modes propagating in the assumed dielectric structure.} By coupling it with the quantum Maxwell's equation,
\begin{align}
	\hat{E}(z, t)= \hat{E}_0(z, t)+\sum_{\mu}\int{\text{d}z'}G(z,z')P_{\mu}(z') \hat{b}_{\mu}(t), \label{Seq:7}
\end{align}
the equations of motion of the exciton system considering self-consistent interactions can then be obtained as follows,
\begin{align}
    \frac{\text{d}\braket{\hat{b}_{\mu}(t)}}{\text{d}t} &= -(i\Omega_{\mu}^{\text{ex}}+\frac{\gamma_{\text{ex}}}{2}+\frac{\Gamma_{\text{ex}}}{2})\braket{\hat{b}_{\mu}(t)} +\frac{i}{\hbar}\int{\text{d}z}P_{\mu}^*(z)\braket{\hat{E}_0(z,t)}-\frac{i}{\hbar}\sum_{\lambda}Z_{\mu \lambda}\braket{\hat{b}_{\lambda}(t)}, \label{Seq:8} 
\end{align}
\begin{align}
    \frac{\text{d}\braket{\hat{b}^{\dagger}_{\mu}(t)\hat{b}_{\mu'}(t)}}{\text{d}t} 
    & = [i(\Omega^{\text{ex}}_{\mu}-\Omega^{\text{ex}}_{\mu'})-\gamma_{\text{ex}}-\Gamma_{\text{ex}}(1-\delta_{\mu,\mu'})]\braket{\hat{b}^{\dagger}_{\mu}(t)\hat{b}_{\mu'}(t)} \nonumber\\
    & + \frac{i}{\hbar}\int{\text{d}z} [\braket{\hat{b}^{\dagger}_{\mu}(t)}P_{\mu'}^*(z)\braket{\hat{E}_0(z,t)} -\braket{\hat{E}^{\dagger}_0(z,t)}P_{\mu}(z)\braket{\hat{b}^{\dagger}_{\mu'}(t)} ] \nonumber\\
    & -\frac{i}{\hbar}\sum_{\lambda}\braket{\hat{b}^{\dagger}_{\mu}(t)\hat{b}_{\lambda}(t)}Z_{\lambda \mu'} + \frac{i}{\hbar}\sum_{\nu}Z_{\mu\nu}^*\braket{\hat{b}^{\dagger}_{\nu}(t)\hat{b}_{\mu'}(t)}, \label{Seq:9}
\end{align}
where $Z_{\mu \mu'}$ is
\begin{eqnarray}
    Z_{\mu \mu'}=-\iint{\text{d}z\text{d}z'}P_{\mu}^*(z)G(z,z'\bla{, \omega_b})P_{\mu'}(z). \label{Seq:10}
\end{eqnarray}
Herein, $Z_{\mu \mu'}$ is the radiative correction term and represents the coupling between excitons through radiation. $G(z,z')$ is the Green's function that considers the structure of a vacuum/SiN/emitter/vacuum/mirror \citep{Chew1995-hr}. \bla{(The analytical expression of this Green's function includes information of all photon modes, $\eta$.)}
In this theory, equations of motion are solved under steady-state conditions. The Fourier transform of equation (\ref{Seq:8}) using the rotating-wave approximation yields
\begin{align}
    &\sum_{\lambda}M_{\mu\lambda}\braket{\hat{b}_{\lambda}(\omega)}=\int{\text{d}z} P_{\mu}^*(z)\braket{\hat{E}_0(z,\omega)}, \label{Seq:11}\\
    &M_{\mu\lambda}(\omega) \equiv \hbar(\Omega^{\text{ex}}_{\mu}-\omega-i\frac{\gamma_{\text{ex}}}{2}-i\frac{\Gamma_{\text{ex}}}{2})\delta_{\mu,\lambda} + Z_{\mu\lambda}, \label{Seq:12}
\end{align}
Solving equation (\ref{Seq:12}) for $N(\omega)=M^{-1}(\omega)$, we obtain 
\begin{eqnarray}
    \braket{\hat{b}_{\mu}(\omega)}=\sum_{\lambda}N_{\mu\lambda}\int{\text{d}z} P_{\lambda}^*(z)\braket{\hat{E}_0(z,\omega)}. \label{Seq:13}
\end{eqnarray}
\bla{Assuming a classical monochromatic continuous wave laser $\braket{\hat{E}_0(z,\omega)}= E_0(z)\delta(\omega-\omega_{\text{in}})$ as the excitation light, the analytical solution under steady state is obtained as follows,}
\begin{eqnarray}
    \braket{\hat{b}_{\mu}}_{\text{ss}} 
    = \sum_{\lambda}N_{\mu\lambda}\int{\text{d}z} P_{\lambda}^*(z)E_0(z).
\label{Seq:14}
\end{eqnarray}
\bla{By substituting the exciton polarization obtained from equation (\ref{Seq:14}) into Maxwell's equation (\ref{Seq:7}), the total electric field, including the radiation field from the exciton polarization, can be described. The calculated optical spectra (reflectance R, transmittance T, and absorption A) were shown in Fig.\ref{Optical_spe_res}. The absorption coefficient under resonant excitation is $A \sim 0.17$, where $\hbar\Gamma_{\text{ex}}=50$ meV, corresponding to the resonant absorption peak in the experiments of Trinh et al. \citep{Trinh2018-ej}.}
%%%%% Figure S1 %%%%%%%
 % CCCCCCCCCCCCCCCCCCCCCC

Using equation (\ref{Seq:14}), we also assume that equations (\ref{Seq:9}) are in the steady state $(\text{d}\braket{\hat{b}^{\dagger}_{\mu}(t)\hat{b}_{\mu'}(t)}/\text{d}t = 0)$, which yields the following,
\begin{align}
    &\sum_{\nu\lambda}S_{\nu\lambda}\braket{\hat{b}^{\dagger}_{\nu}\hat{b}_{\lambda}}_{\text{ss}} 
    =-\int{\text{d}z} [\braket{\hat{b}^{\dagger}_{\mu}}_{\text{ss}}P_{\mu'}^*(z)E_0(z)-E_0^*(z)P_{\mu}(z)\braket{\hat{b}^{\dagger}_{\mu'}}_{\text{ss}}],  \label{Seq:15} \\
    S_{\nu\lambda} &= [\{ \hbar(\Omega^{\text{ex}}_{\mu}-\Omega^{\text{ex}}_{\mu'})+i\hbar\gamma_{\text{ex}} +i\hbar\Gamma_{\text{ex}}(1-\delta_{\mu,\mu'}) \}\delta_{\lambda,\mu'}\delta_{\nu,\mu'} - Z_{\lambda\mu'} + Z_{\mu\nu}^*]. \label{Seq:16}
\end{align}
By solving the above equation, we can obtain the number of excitons $\braket{\hat{b}^{\dagger}_{\mu}\hat{b}_{\mu}}_{\text{ss}}$ and the correlations $\braket{\hat{b}^{\dagger}_{\mu}\hat{b}_{\mu'}}_{\text{ss}}$ at each excitonic level.

\begin{figure}[htbp] % CCCCCCCCCCCCCCCCCCCCCC
        \centering
          \includegraphics[width=90mm]{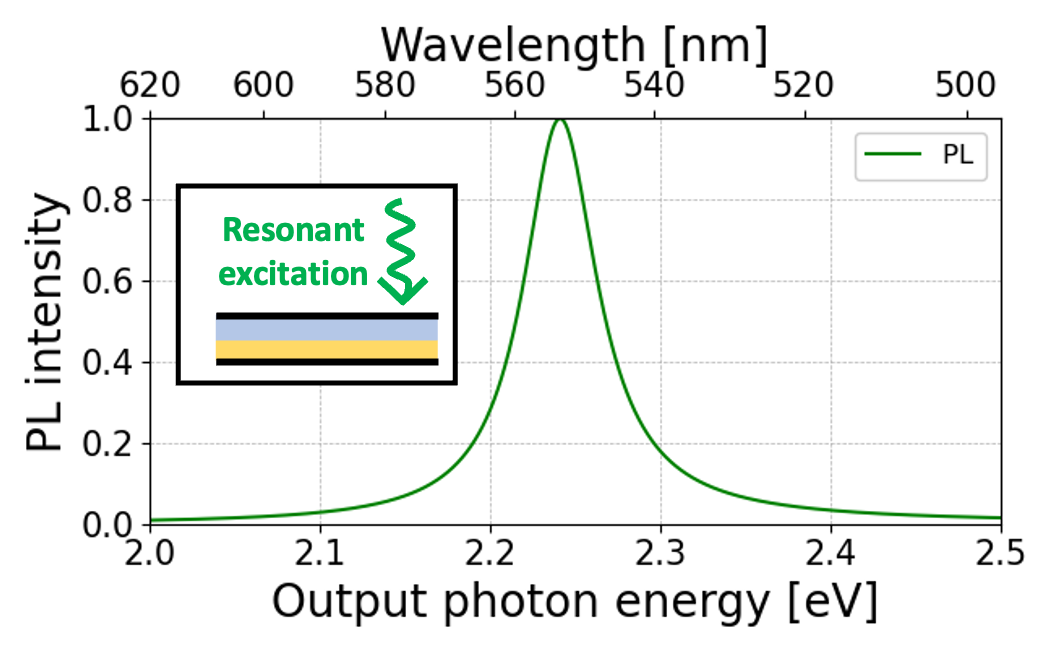}
    \caption{
    \bla{Resonance} photoluminescence (PL) spectrum of the composite film of SiN with thickness $w = 50$ nm and $\bla{\text{FAPbBr}_{3}}$ layer with thickness $d = 50$ nm. The excitation light energy $\hbar\omega_{\text{in}}$ was set in this study to to the transverse-wave exciton energy $\hbar\omega_{\text{in}}=\hbar\omega_{\bla{b}}\bla{=2.24}$ \bla{eV}. The PL spectrum data is normalized by peak intensity. \bla{The FWHM is around 12 nm, which is derived from experimental works of Hu et al. \citep{Hu2022-zt}. }
        \label{figS2}
        }
\end{figure}

The \bla{resonance} PL theory is described next. Herein, the electric field intensity of the incoherent light, which is the total response electric field intensity minus the coherent electric field intensity, is treated as the PL intensity and the corresponding PL spectrum $S_{\text{inc}}(\omega)$ is expressed as follows,
\begin{align}
    &S_{\text{inc}}(z,\omega) = \frac{1}{\pi}\operatorname{Re}\bigg[ \sum_{\mu,\mu'}F_{\mu}^*(z)F_{\mu'}(z) \Delta B_{\mu\mu'}(\omega) \bigg],\label{Seq:17}\\
    &\Delta B_{\mu\mu'}(\omega) = \int_0^{\infty}{\mathrm{d}\tau}\braket{\Delta\hat{b}_{\mu}^{\dagger}(t)\Delta\hat{b}_{\mu'}(t+\tau)} \mathrm{e}^{i\omega\tau}, \label{Seq:18}\\
    &\braket{\Delta\hat{b}_{\mu}^{\dagger}(t)\Delta\hat{b}_{\mu'}(t+\tau)} \equiv \braket{\hat{b}_{\mu}^{\dagger}(t)\hat{b}_{\mu'}(t+\tau)} - \braket{\hat{b}_{\mu}^{\dagger}(t)}\braket{\hat{b}_{\mu'}(t+\tau)},\label{Seq:19}
\end{align}
where $F_{\mu}(z) = \int{\mathrm{d}z'}G(z,z',\omega)P_{\mu}(z')$. 
The Heisenberg equation for the first-order correlation function $\braket{\Delta\hat{b}_{\mu}^{\dagger}(0)\Delta\hat{b}_{\mu'}(\tau)}$ is expressed using the quantum regression theorem as follows,
\begin{align}
    &\frac{\text{d}\braket{\Delta\hat{b}_{\mu}^{\dagger}(0)\Delta\hat{b}_{\mu'}(\tau)}}{\text{d}\tau}
    = -(i\Omega_{\mu'}^{\text{ex}}+\frac{\gamma_{\text{ex}}}{2}+\frac{\Gamma_{\text{ex}}}{2})\braket{\Delta\hat{b}_{\mu}^{\dagger}(0)\Delta\hat{b}_{\mu'}(\tau)}
    -\frac{i}{\hbar}\sum_{\lambda}Z_{\mu' \lambda}\braket{\Delta\hat{b}_{\mu}^{\dagger}(0)\Delta\hat{b}_{\lambda}(\tau)}. \label{Seq:20}
\end{align}
\bla{The emission linewidth is essentially determined by phase relaxation constant $\Gamma_{\text{ex}}$. The FWHM of the PL spectrum of $\text{FAPbBr}_3$ at cryogenic temperatures has been experimentally measured to be approximately 12 nm \citep{Hu2022-zt}. Accordingly, $\hbar\Gamma_{\text{ex}} = 50$ meV is used in this paper.} The Fourier transform of Eq. (\ref{Seq:20}),
\begin{align}
    \sum_{\lambda}M_{\mu'\lambda}\Delta B_{\mu\lambda}(\omega)=-i\hbar\braket{\Delta\hat{b}_{\mu}^{\dagger}(0)\Delta\hat{b}_{\mu'}(0)}, \label{Seq:21}
\end{align}
can be solved as follows,
\begin{align}
    &\Delta B_{\mu\mu'}(\omega)=-i\hbar\sum_{\mu'}N_{\lambda\mu'}\braket{\Delta\hat{b}_{\mu}^{\dagger}(0)\Delta\hat{b}_{\lambda}(0)}, \label{Seq:22}\\
    &\braket{\Delta\hat{b}_{\mu}^{\dagger}(0)\Delta\hat{b}_{\mu'}(0)} = \braket{\hat{b}_{\mu}^{\dagger}\hat{b}_{\mu'}}_{\text{ss}} - \braket{\hat{b}_{\mu}^{\dagger}}_{\text{ss}}\braket{\hat{b}_{\mu'}}_{\text{ss}}. \label{Seq:23}
\end{align}
The calculated PL spectrum is shown in Fig. \ref{figS2}.

%%%%%%% Figure S2 %%%%%%%
 % CCCCCCCCCCCCCCCCCCCCCC

%%%%%%%%%%%% Appendix %%%%%%%%%%%%
\section{PL theory for luminescent film under non-resonant excitation}
\begin{figure}[htbp] % CCCCCCCCCCCCCCCCCCCCCC
        \centering
          \includegraphics[width=90mm]{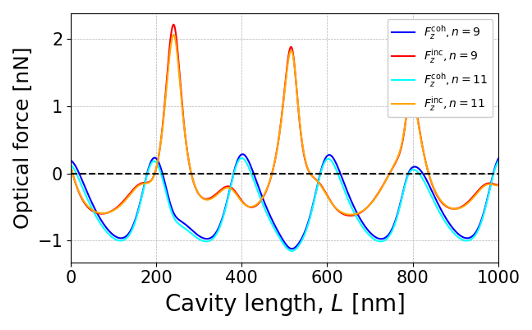}
    \caption{
    \bla{Dependence of the (Red) \bla{LIOF} and (Blue) optical force induced by coherent light on the cavity length $L$ when the number of exciton states $N_b$ and $N_c$ were varied to 9 and 11. The assumptions and parameters are identical to those described in Figure 3(a) of the main text.}
        \label{L-OF_n=9or11}
        }
\end{figure}

\bla{In Section \ref{Sec:Res_PL}, we explained the exciton-radiation field interaction under resonant excitation, as well as the resonance PL theory. 
In this section, we developed a theoretical framework to describe the PL from the emissive levels $\hat{b}$ under excitation by light with wavelengths shorter than the emission band. This is modeled by introducing energy levels (excited states $\hat{c}$) that are resonant with the excitation light, allowing resonant pumping of these levels, followed by nonradiative relaxation to the emissive levels $\hat{b}$. This framework accounts for non-resonant emission while incorporating the effects of the Stokes shift. The Hamiltonian is modified to include the modes of the excited states, as shown in Eq. (\ref{Seq:1}):
\begin{align}
    \hat{H} =& \sum_{\eta} \hbar\Omega_{\eta} \hat{a}^{\dagger}_{\eta} \hat{a}_{\eta}
    + \sum_{o}^{b,c}\sum_{\mu=1}^{N_{o}} \hbar \Omega_{\mu, o}^{\text{ex}} \hat{o}_\mu^\dagger \hat{o}_\mu
    -\sum_{o}^{b,c} \int{\text{d} z} \hat{P}_{\text{ex}, o}(z) \hat{E}(z),
\end{align}
where $\hat{o}^{\dagger}_{\mu} (\hat{o}_{\mu})$ represents the creation (annihilation) operator of the $\mu$th exciton state. We assumed two modes for the exciton: emission levels $\hat{b}$ and an excitation-levels $\hat{c}$. $\hat{P}_{\text{ex},o}(z)=\sum_{\mu} (P_{\mu,o}(z)\hat{o}_{\mu}(t) + \text{H.c.}$) is the excitonic polarization operator. Let $P_{\mu,o}(z)$ be expressed as $P_{\mu,o}(z)=P_{o}\psi_{\mu}(z)$ using $P_{o}^{2} = \varepsilon_0\varepsilon_{\mathrm{b}}\Delta_{\text{LT},o}$ and let the wave function of excitonic center-of-mass motion $\psi_{\mu}(z)$ be expressed as $\psi_{\mu}(z) = \sqrt{2/d}\sin{K_{\mu}z}$ in quantized form with the $\sin$ function ($K_{\mu} = \mu\pi/d, \mu = 1,2,\cdots, N_o$). The exciton eigenenergy $\hbar\Omega_{\mu,o}^{\text{ex}}$ can also be expressed as $\hbar\Omega_{\mu,o}^{\text{ex}} = \hbar\omega_{o} + \hbar^2 K_{\mu}^2/(2m_{\text{ex}})$.
Here, $\mu$ represents the order of exciton modes. In our calculations, we considered up to the 10th mode ($N_b=N_c=10$), deemed sufficient given the small thickness of the luminescent film relative to the wavelength of light. This consideration is based on the understanding that the states interacting with the optical field are predominantly the lower-order modes ($\mu \simeq 1,2$). Figure \ref{L-OF_n=9or11} shows the optical force when the number of exciton states is slightly increased or decreased ($N_b=N_c=9, 11$). It can be observed that varying the number of modes considered does not significantly affect the results for the optical force.} 

%%%%%%% Figure S3 %%%%%%%
 % CCCCCCCCCCCCCCCCCCCCCC

\begin{figure}[htbp] % CCCCCCCCCCCCCCCCCCCCCC
        \centering
          \includegraphics[width=90mm]{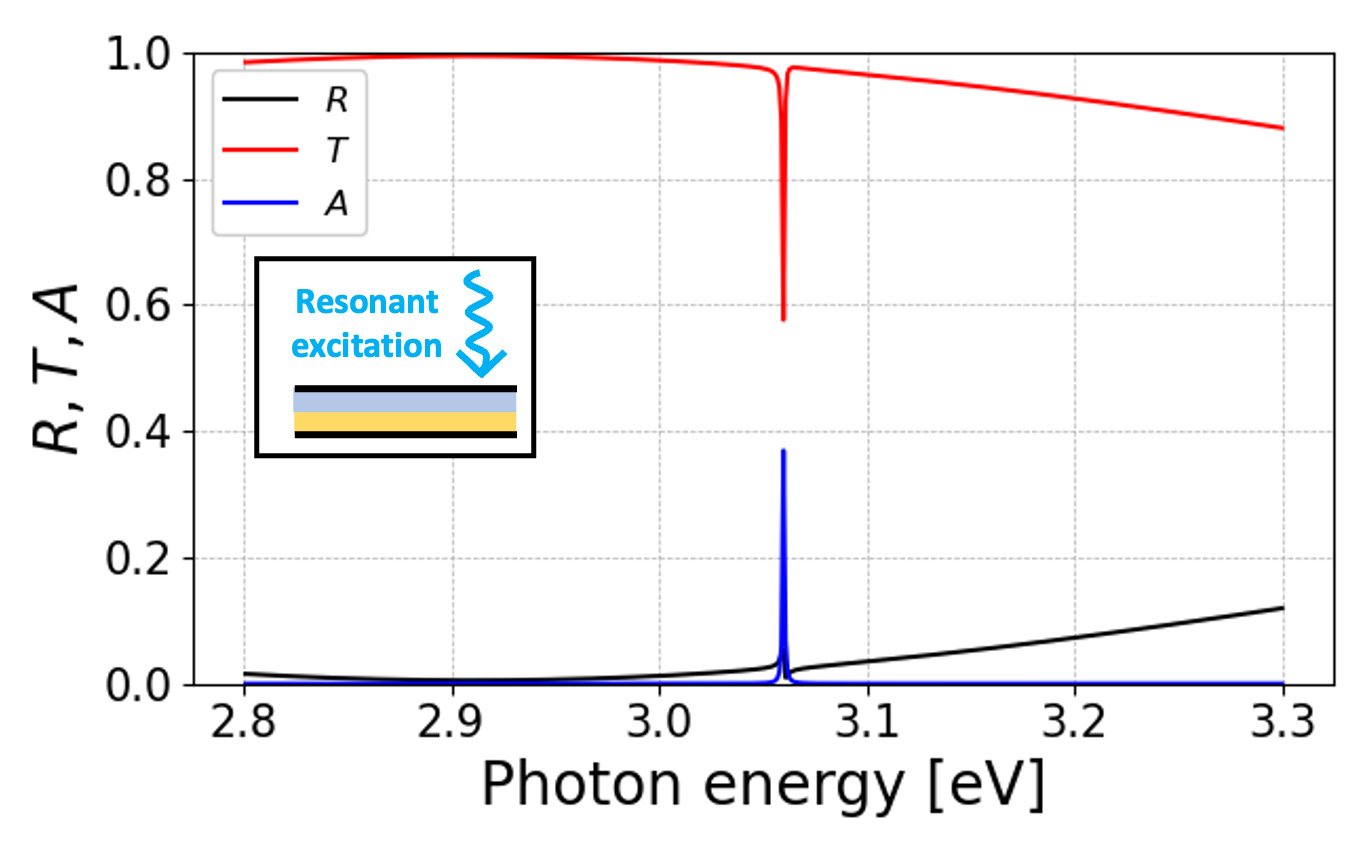}
    \caption{
    \bla{Reflection (black line: R), transmission (red line: T), and absorption (blue line: A) spectra under non-resonant excitation. The composite film consists of SiN and $\text{FAPbBr}_{3}$ layer with thickness $w = d = 50$ nm. The black dotted line represents the excitation light energy, which is set to $\hbar\omega_{\text{in}}=2.24$ eV in this paper. Herein, $\hbar\Gamma_{c}=0.4$ meV and $\Delta_{\text{LT},c} = 0.03\Delta_{\text{LT},b}$. The absorption coefficient under 405 nm excitation is $A \sim 0.37$, corresponding to the absorption ratio in the experiments of Trinh et al. \citep{Trinh2018-ej}.}
        \label{Optical_spe_nonres}
        }
\end{figure}

\bla{The quantum master equation is expressed as follows by adding inter-level nonradiative damping processes, 
\begin{align}
	\frac{\partial \hat{\rho}(t)}{\partial t}
        &=\frac{1}{i\hbar}[\hat{H},\hat{\rho}(t)] + \sum_{o}^{b,c}\sum_{\mu} \frac{\gamma_{\text{o}, \mu}}{2}\hat{L}_{\hat{o}_{\mu}}\hat{\rho}(t) + \sum_{\mu,\mu'} \frac{\gamma_{\text{c-b}}^{\mu,\mu'}}{2}\hat{L}_{\hat{b}^{\dagger}_{\mu'}\hat{c}_{\mu}}\hat{\rho}(t) + \sum_{o}^{b,c}\frac{\Gamma_{\text{o}}}{2}\hat{L}_{\text{phase}}(o)\hat{\rho}(t), \label{Aeq:2}\\
        &\hat{L}_{\text{damp},\hat{O}}\hat{\rho}(t) = [2\hat{O}\hat{\rho}(t)\hat{O}^{\dagger} - \{\hat{O}^{\dagger}\hat{O},\hat{\rho}(t)\} ], \label{Aeq:3}\\
        &\hat{L}_{\text{phase}}(o)\hat{\rho}(t) = \sum_{\mu}\big[[\hat{o}^{\dagger}_{\mu}\hat{o}_{\mu},\hat{\rho}(t)],\hat{o}^{\dagger}_{\mu}\hat{o}_{\mu} \big], 
        % \label{Aeq:4}
\end{align}
where $\hat{O}$ is annihilation operator, which includes $\hat{O}=\hat{b}_{\mu}, \hat{b}^{\dagger}_{\mu'}\hat{c}_{\mu}$. The second term on the right-hand side of Eq. (\ref{Aeq:2}) represents the nonradiative damping of each emission levels, and the third term represents the inter-level damping from the excited level ($\hat{c}_{\mu}$) to the emission levels ($\hat{b}_{\mu'}$). The fourth represents the phase relaxation term. The nonradiative damping constants are assumed to be the same for all levels ($\gamma_{\text{o}, \mu}=\gamma_{\text{o}}$), and the decay constants are $\hbar\gamma_{b} 
 = \hbar\gamma_{c} = 0.2$ \textmu eV \citep{He2018-nd}. The inter-level damping is assumed to be determined in accordance with the radiative damping of the radiative levels, $\hbar\gamma_{c-b}^{\mu,\mu'} = -2\Sigma_{\mu'}\operatorname{Im}[Z_{\mu'\mu',b}]/N_c$. The dephasing constant $\hbar\Gamma_{c}=0.4$ meV \citep{Tamarat2019-tk}. From Equation (\ref{Aeq:2}), the Heisenberg equation for the expected value of each exciton operator can be obtained as follows by coupling it with the quantum Maxwell's equation, $\hat{E}(z, t)= \hat{E}_0(z, t)+ \sum_{o}^{b,c}\sum_{\mu}\int{\text{d}z'}G(z,z', \omega_{o})P_{\mu,o}(z') \hat{o}_{\mu}(t)$, 
\begin{align}
    \frac{\text{d}\braket{\hat{o}_{\mu}(t)}}{\text{d}t} &= -\left[i\Omega_{\mu}^{\text{ex,o}}+\frac{\gamma_{\text{o}}}{2}+\frac{\Sigma_{\mu'}\gamma_{\text{c-b}}^{\mu,\mu'}}{2}\delta_{o,c}+\frac{\Gamma_{o}}{2}\right]\braket{\hat{o}_{\mu}(t)} +\frac{i}{\hbar}\int{\text{d}z}P_{\mu,o}^{*}(z)\braket{\hat{E}_0(z,t)}-\frac{i}{\hbar}\sum_{\lambda}Z_{\mu \lambda,o}\braket{\hat{o}_{\lambda}(t)}, 
    \label{Seq:28} 
\end{align}
\begin{align}
    \frac{\text{d}\braket{\hat{o}^{\dagger}_{\mu}(t)\hat{o}_{\mu'}(t)}}{\text{d}t} 
    & = [i(\Omega^{\text{ex}}_{\mu, o}-\Omega^{\text{ex}}_{\mu', o})-\gamma_{\text{o}}(1-\delta_{o,\mu}\delta_{o,\mu'})-\Sigma_{\mu'}\gamma_{\text{c-b}}^{\mu,\mu'}(1-\delta_{o,\mu}\delta_{o,\mu'})-\Gamma_{\text{o}}(1-\delta_{\mu,\mu'})]\braket{\hat{o}^{\dagger}_{\mu}(t)\hat{o}_{\mu'}(t)} \nonumber\\
    & + \frac{i}{\hbar}\int{\text{d}z} [\braket{\hat{o}^{\dagger}_{\mu}(t)}P_{\mu',o}^{*}(z)\braket{\hat{E}_0(z,t)} -\braket{\hat{E}^{\dagger}_0(z,t)}P_{\mu,o}(z)\braket{\hat{o}^{\dagger}_{\mu'}(t)} ] \nonumber\\
    & -\frac{i}{\hbar}\sum_{\lambda}\braket{\hat{o}^{\dagger}_{\mu}(t)\hat{o}_{\lambda}(t)}Z_{\lambda \mu', o} + \frac{i}{\hbar}\sum_{\nu}Z_{\mu\nu, o}^*\braket{\hat{o}^{\dagger}_{\nu}(t)\hat{o}_{\mu'}(t)}, 
    \label{Seq:29}
\end{align}
where $Z_{\mu \mu',o}$ is
\begin{eqnarray}
    Z_{\mu \mu',o}=-\iint{\text{d}z\text{d}z'}P_{\mu,o}^{*}(z)G(z,z',\omega_o)P_{\mu',o}(z). 
    % \label{Aeq:8}
\end{eqnarray}
Herein, $Z_{\mu \mu', o}$ is the radiative correction term and represents the coupling between excitons through radiation. $G(z,z',\omega_o)$ is Green's function, and these equations of motion of the exciton system consider self-consistent interactions through radiation. Here, the energy levels of the excited state and the emissive state are assumed to be sufficiently separated, such that no interaction via radiation occurs between them. Additionally, the self-interactions of the emissive and excited states are calculated under the assumption that they are radiatively coupled through the Green's function at each resonant frequency $\omega_o$.
Solving equations of motion under steady-state conditions, the Fourier transform of equation (\ref{Seq:28}) using the rotating-wave approximation yields
\begin{align}
    &\sum_{\lambda}M_{\mu\lambda,o}\braket{\hat{o}_{\lambda}(\omega)}=\int{\text{d}z} P_{\mu,o}^{*}(z)\braket{\hat{E}_0(z,\omega)}, 
    \label{Seq:31}
    \\
    &M_{\mu\lambda,o}(\omega) \equiv \hbar(\Omega^{\text{ex}}_{\mu,o}-\omega-i\frac{\gamma_{\text{o}}}{2}-i\frac{\Gamma_{\text{o}}}{2})\delta_{\mu,\lambda} + Z_{\mu\lambda,o}, 
    \label{Seq:32}
\end{align}
Solving equation (\ref{Seq:31}) for $N(\omega)=M^{-1}(\omega)$, we obtain the analytical solution:
\begin{eqnarray}
    \braket{\hat{o}_{\mu}}_{\text{ss}} 
    = \braket{\hat{o}_{\mu}(\omega)}=\sum_{\lambda}N_{\mu\lambda,o}\int{\text{d}z} P_{\lambda,o}^{*}(z)\braket{\hat{E}_0(z,\omega)}, 
    % \label{Aeq:11}
\end{eqnarray}
and can be obtained under steady-state conditions by assuming a classical monochromatic continuous wave laser $(\braket{\hat{E}_0(z,\omega)}= E_0(z)\delta(\omega-\omega_{\text{in}})$ as the excitation light. The calculated optical spectra were shown in Fig.\ref{Optical_spe_nonres}. The absorption coefficient under 405 nm excitation is $A \sim 0.37$, corresponding to the absorption ratio in the experiments of Trinh et al. \citep{Trinh2018-ej}.} 
%%%%%%% Figure %%%%%%%
 % CCCCCCCCCCCCCCCCCCCCCC

\bla{The equations are follows,
\begin{align}
    &\sum_{\nu,\lambda}S_{\nu\lambda,o}\braket{\hat{o}^{\dagger}_{\nu}\hat{o}_{\lambda}}_{\text{ss}} 
    =-\int{\text{d}z} [\braket{\hat{o}^{\dagger}_{\mu}}_{\text{ss}}P_{\mu',o}^*(z)E_0(z)-E_0^*(z)P_{\mu,o}(z)\braket{\hat{o}^{\dagger}_{\mu'}}_{\text{ss}}],  
    % \label{Aeq:13}
\end{align}
\begin{align}
    S_{\nu\lambda,o} &= [\{ \hbar(\Omega^{\text{ex}}_{\mu,o}-\Omega^{\text{ex}}_{\mu',o})+i\hbar\gamma_{o} +i\hbar\Gamma_{o}(1-\delta_{\mu,\mu'}) \}\delta_{\lambda,\mu'}\delta_{\nu,\mu'} - Z_{\lambda\mu',o} + Z_{\mu\nu,o}^*]. 
    % \label{Aeq:14}
\end{align}
By solving the above equation, we can obtain the number of excitons $\braket{\hat{o}^{\dagger}_{\mu}\hat{o}_{\mu}}_{\text{ss}}$ and the correlations $\braket{\hat{o}^{\dagger}_{\mu}\hat{o}_{\mu'}}_{\text{ss}}$ at each excitonic level.}

\begin{figure}[htbp] % CCCCCCCCCCCCCCCCCCCCCC
        \centering
          \includegraphics[width=90mm]{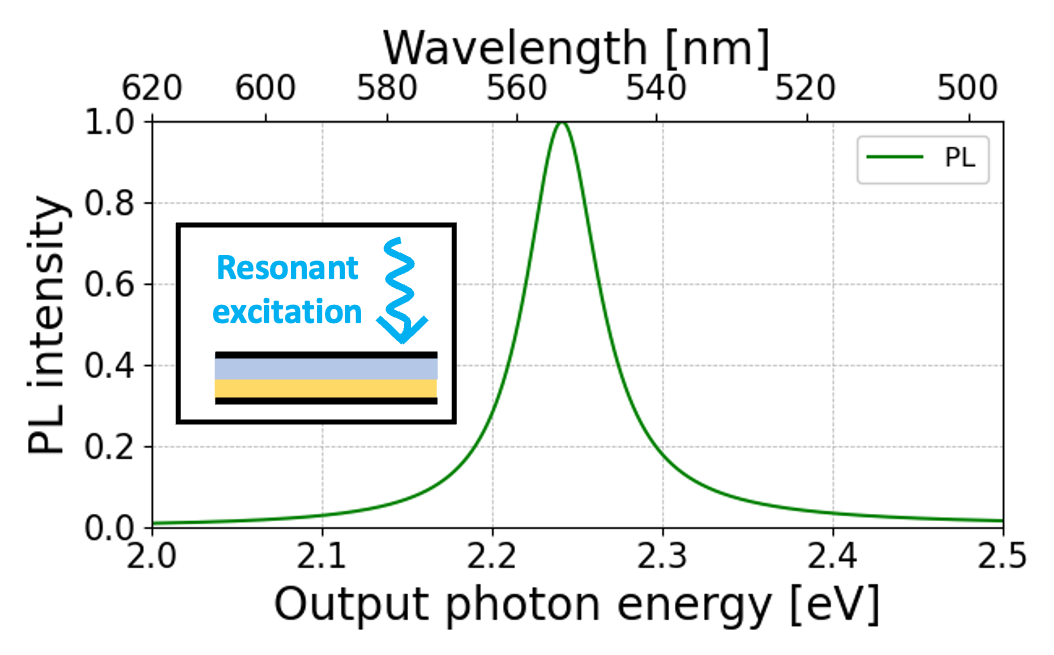}
    \caption{
    \bla{PL spectrum of the composite film (SiN, $w = 50$ nm and $\text{FAPbBr}_{3}$ layer, $d = 50$ nm) under non-resonant excitation. The excitation light energy $\hbar\omega_{\text{in}}$ was set $\hbar\omega_{\text{in}}=3.06$ eV. The PL spectrum data is normalized by peak intensity. Herein, $\hbar\Gamma_{b}=50$ meV, $\hbar\Gamma_{c}=0.4$ meV and $\Delta_{\text{LT},c} = 0.03\Delta_{\text{LT},b}$. The FWHM is around 12 nm, which is derived from experimental works of Hu et al. \citep{Hu2022-zt}. }
        \label{PL_nonres}
        }
\end{figure}

\bla{The PL spectrum $S_{\text{inc}}(\omega)$ is expressed as follows,
\begin{align}
    &S_{\text{inc}}(z,\omega) = \frac{1}{\pi}\operatorname{Re}\bigg[ \sum_{\mu,\mu'}F_{\mu}^*(z)F_{\mu'}(z) B_{\mu\mu'}(\omega) \bigg],\label{Seq:36}\\
    & B_{\mu\mu'}(\omega) = \int_0^{\infty}{\mathrm{d}\tau}\braket{\hat{b}_{\mu}^{\dagger}(t)\hat{b}_{\mu'}(t+\tau)} \mathrm{e}^{i\omega\tau}, \label{Seq:37}
\end{align}
where $F_{\mu}(z) = \int{\mathrm{d}z'}G_b(z,z',\omega)P_{\mu,b}(z')$. 
The Heisenberg equation for the first-order correlation function $\braket{\hat{b}_{\mu}^{\dagger}(0)\hat{b}_{\mu'}(\tau)}$ is expressed using the quantum regression theorem as follows,
\begin{align}
    &\frac{\text{d}\braket{\hat{b}_{\mu}^{\dagger}(0)\hat{b}_{\mu'}(\tau)}}{\text{d}\tau}
    = -(i\Omega_{\mu'}^{\text{ex}}+\frac{\gamma_{\text{e-g}}}{2}+\frac{\Gamma_{\text{c}}}{2})\braket{\hat{b}_{\mu}^{\dagger}(0)\hat{b}_{\mu'}(\tau)}
    -\frac{i}{\hbar}\sum_{\lambda}Z_{\mu' \lambda}\braket{\hat{b}_{\mu}^{\dagger}(0)\hat{b}_{\lambda}(\tau)}. \label{Seq:38}
\end{align}
The Fourier transform of Eq. (\ref{Seq:38}),
\begin{align}
    \sum_{\lambda}M_{\mu'\lambda} B_{\mu\lambda}(\omega)=-i\hbar\braket{\hat{b}_{\mu}^{\dagger}(0)\hat{b}_{\mu'}(0)}, \label{Seq:39}
\end{align}
can be solved as follows,
\begin{align}
    & B_{\mu\mu'}(\omega)=-i\hbar\sum_{\mu'}N_{\lambda\mu'}\braket{\hat{b}_{\mu}^{\dagger}(0)\hat{b}_{\lambda}(0)}, \label{Seq:40}
\end{align}
The calculated PL spectrum is shown in Fig.\ref{PL_nonres}.}

%%%%%%% Figure %%%%%%%
 % CCCCCCCCCCCCCCCCCCCCCC

%%%%%%%%%%%% Appendix %%%%%%%%%%%%
\section{LIOF theory for luminescent films}\label{Sec:LIOF}
The \bla{LIOF} $\ev{f_z^{\text{inc}}(\omega)}$ generated in the direction perpendicular to the film surface by light at the emission frequency $\omega$ can be derived from Maxwell's stress tensor (see equation below), where $\ev{}$ is the time-averaged value over $\omega$ within the light oscillation period,
\begin{align}
    &\left\langle f_z^{\mathrm{inc}}(\omega, L)\right\rangle 
    =\frac{S}{2} \varepsilon_0\left(\left|E_{\mathrm{L} 1}(\omega, L)\right|^2+\left|E_{\mathrm{L} 2}(\omega, L)\right|^2 -\left|E_{\mathrm{U} 1}(\omega, L)\right|^2\right),  \label{Deq:1}
\end{align}
where $S$ is the excited area of the luminescent film, $\varepsilon_0$ is the dielectric constant of vacuum, $\text{E}_\text{L1} (\text{E}_{\text{L2}})$ denotes the optical field of the upward (downward) wavenumber at the lower surface of the film in the model in Fig. 1 in the main text, and $\text{E}_{\text{U1}}$ represents the optical field of the upward wavenumber at the upper surface of the film. The luminescence intensity at the film surface is expressed based on the equation listed below, where $k_z$ is the emission wavenumber in the $z$ direction,
\begin{align}
    &S_{\mathrm{inc}}(z_{\bla{\text{btm}}}=L, \omega) 
    =\left|E_{\mathrm{L} 1} e^{i k_z L}+E_{\mathrm{L} 2} e^{-i k_z L}\right|^2 \nonumber\\
    & \quad =\left|E_{\mathrm{L} 1}\right|^2+\left|E_{\mathrm{L} 2}\right|^2+2 \operatorname{Re}\left\{\left[E_{\mathrm{L} 1} E_{\mathrm{L} 2}\right]\right\} \operatorname{Re}\left\{\left[e^{2 i k_z L}\right]\right\} -2 \operatorname{Im}\left\{\left[E_{\mathrm{L} 1} E_{\mathrm{L} 2}\right]\right\} \operatorname{Im}\left\{\left[e^{2 i k_z L}\right]\right\}, \label{Deq:2}\\
    &S_{\mathrm{inc}}(z_{\bla{\text{up}}}=L+d+w, \omega) =\left|E_{\mathrm{U} 1}(\omega)\right|^2.
\end{align}
To calculate \bla{LIOF} using Eq. (\ref{Deq:1}), we require the sum of the squares of the electric field amplitudes $\left|E_{\mathrm{L} 1}(\omega)\right|^2+\left|E_{\mathrm{L} 2}(\omega)\right|^2$ for each emission wavenumber. This sum is obtained by calculating the luminescence intensity at three arbitrary positions $(z_1, z_2, z_3)$ that differ from each other inside the cavity, and by solving the simultaneous equation based on Eq. (\ref{Deq:2}).
\begin{align}
    & \left(\begin{array}{ccc}
    1 & 2 \operatorname{Re}\left[\mathrm{e}^{2 i k_z z_1}\right] & 2 \operatorname{Im}\left[\mathrm{e}^{2 i k_z z_1}\right] \\
    1 & 2 \operatorname{Re}\left[\mathrm{e}^{2 i k_z z_2}\right] & 2 \operatorname{Im}\left[\mathrm{e}^{2 i k_z z_2}\right] \\
    1 & 2 \operatorname{Re}\left[\mathrm{e}^{2 i k_z z_3}\right] & 2 \operatorname{Im}\left[\mathrm{e}^{2 i k_z z_3}\right]
    \end{array}\right)
    \left(\begin{array}{c}
    \left|E_{\mathrm{L} 1}\right|^2+\left|E_{\mathrm{L} 2}\right|^2 \\
    \operatorname{Re}\left[E_{\mathrm{L} 1} E_{\mathrm{L} 2}^*\right] \\
    \operatorname{Im}\left[E_{\mathrm{L} 1} E_{\mathrm{L} 2}^*\right]
    \end{array}\right) 
    = \left(\begin{array}{c}
    S_{\mathrm{inc}}\left(z=z_1, \omega\right) \\
    S_{\mathrm{inc}}\left(z=z_2, \omega\right) \\
    S_{\text {inc }}\left(z=z_3, \omega\right)
\end{array}\right).
\end{align}
Finally, \bla{LIOF} $\ev{F_z^{\text{inc}}}$ is obtained by integrating $\ev{f_z^{\text{inc}}}$ at the emission frequency $\omega$ as follows,
\begin{align}
    \left\langle F_z^{\mathrm{inc}}(L)\right\rangle=\int\left\langle f_z^{\text {inc }}(\omega, L)\right\rangle \mathrm{d} \omega.
\end{align}

%%%%%%%%%%%% Appendix E %%%%%%%%%%%%
\section{Equation of motion for square-type membrane}\label{Sec:Mech_mem}
We assumed a rectangular membrane with lengths $a$ and $b$ per side and thickness $d$. A two-dimensional wave equation is expressed as follows,
\begin{align}
    \frac{\mathrm{d}^2}{\mathrm{~d} t^2} A(x, y, t) -\frac{T_{\text{m}}}{\rho} \Delta_{x, y} A(x, y, t) = 0
\end{align}
where $A$ denotes the vibration amplitude at position $(x,y)$, $T_{\text{m}}$ denotes the membrane tension, $\rho$ denotes the membrane mass density, and $\Delta_{x, y}$ denotes the two-dimensional Laplacian. Using the boundary condition $A(x,y) = 0$, the vibration amplitude of a rectangular membrane for modes $(m, n)$ is expressed as follows, \citep{Wilson2009-jn,Liu2011-rb}:
\begin{align}
    A_{m, n}(x, y, t) &= A_0 \sin \frac{m \pi x}{a} \sin \frac{n \pi y}{b} \cos \omega_{m, n} t, \\
    \omega_{m, n} &= \sqrt{\frac{T_{\text{m}}}{\rho}\left\{\left(\frac{m \pi}{a}\right)^2+\left(\frac{n \pi}{b}\right)^2\right\}},
\end{align}
where $A_0$ denotes the maximum amplitude. Considering the kinetic energy per unit volume,
\begin{align}
    K(x, y, z) =\frac{1}{2}\rho \ev{\frac{\mathrm{d} A_{m, n}(x, y, t)}{{\mathrm{d} t}}}^2,
\end{align}
The kinetic energy of the entire membrane is,
\begin{align}
    & \int_0^a \int_0^b \int_0^d K(x, y, z) \mathrm{d}x \mathrm{d}y \mathrm{d} z =\frac{1}{2} \times \frac{1}{4} \rho a b d \omega_{m, n}^2 \times\left(\frac{A_0}{\sqrt{2}}\right)^2=\frac{1}{2} m_{\mathrm{eff}} \omega_{m, n}^2 \times\left(\frac{A_0}{\sqrt{2}}\right)^2.
\end{align}
Based on the considerations listed above, a square membrane oscillator with an effective mass $m_{\text{eff}} = m/4$ can be 
regarded as a one-dimensional damped harmonic oscillator vibrating at a frequency $\omega_{m, n}$ with an average amplitude $A_0/\sqrt{2}$, where the membrane's physical mass is $m = \rho abd$.
Thus, the mechanical motion of the film can be treated using the following equation of motion \citep{Duy-Vy2013-th},
\begin{align}
    m_{\text{eff}}\frac{\mathrm{d}^2L(t)}{\mathrm{d}t^2} 
    &+ m_{\text{eff}}\Gamma_{\text{m}}\frac{\mathrm{d}L(t)}{\mathrm{d}t} + m_{\text{eff}}\omega_{\text{m}}^2(L(t)-L_{\text{ini}}) 
    = \ev{F_{z}(L)} + F_z^{\text{th}}(t) + F_z^{\text{act}}(t), \label{Eeq:6}
\end{align}
where $L_{\text{ini}}$ is the initial position of the film, $F_z^{\text{th}}$ is the external noise, $\ev{F_z^{\text{th}}(t)F_z^{\text{th}}(t')} = 2m_{\text{eff}} \Gamma_{\text{m}} k_B T\delta(t-t')$, and $T$ is the temperature. $m_{\text{eff}}$ denotes the effective mass when the oscillator vibration mode is considered. The thicknesses of the $\bla{\text{FAPbBr}_{3}}$ film and SiN membrane are denoted as $d$ and $w$, respectively. Their corresponding densities are $\rho_{\bla{\text{FAPbBr}_{3}}}=\bla{3.67}$ $\text{g}/\text{cm}^3$ for $\bla{\text{FAPbBr}_{3}}$ and $\rho_{\text{SiN}}=2.80$ $\text{g}/\text{cm}^3$ for SiN. The effective mass of the oscillator is as follows,
\begin{align}
    m_{\text{eff}} = \frac{1}{4}(\rho_{\bla{\text{FAPbBr}_{3}}}S_{\bla{\text{FAPbBr}_{3}}}d + \rho_{\text{SiN}}S_{\text{SiN}}w), \label{Eeq:7}
\end{align}
where $S_{\bla{\text{FAPbBr}_{3}}}$($S_{\text{SiN}}$) is the area of $\bla{\text{FAPbBr}_{3}}$ film (SiN membrane). Furthermore, $\omega_{\text{m}}$ is the mechanical frequency of the resonator, and $\Gamma_{\text{m}}$ is the mechanical damping constant obtained by $\Gamma_{\text{m}}=\omega_{\text{m}}/Q_{\text{m}}$ using the mechanical quality factor $Q_{\text{m}}$ of the resonator. In this study, calculations were performed with $\omega_{\text{m}}/2\pi = \bla{400}$ \bla{k}Hz, $Q_{\text{m}} = \bla{3 \times 10^4}$, and $T = 4$ K. The thermal noise at a small time $\Delta t$ is considered as $F_z^{\text{th}} (t) = \sqrt{2m_{\text{eff}} \Gamma_{\text{m}} k_B T} \xi \Delta t^{-1/2}$ \citep{Hoffmann2010-dj} where $\xi$ is a Gaussian random number. The force induced by the actuation laser was assumed to be described by $F_{z}^{\text{act}} = F_{z,0}^{\text{act}} \cos{\omega t}$.

The power spectral density $S_{\text{m}}(\omega)$ is expressed by performing a Fourier transformation of Eq.(\ref{Eeq:6}).
\begin{align}
    \tilde{L}(\omega) &= \frac{1}{m_{\mathrm{eff}}} \frac{F_{z,0}^{\mathrm{act}}}{\omega_{\mathrm{eff}}^2-\omega^2+i \omega \Gamma_{\mathrm{eff}}} \\
    S_{\mathrm{m}}(\omega) &= \abs{\tilde{L}(\omega)}^2 = \frac{1}{m_{\mathrm{eff}}^2} \frac{\left|F_{z, 0}^{\text{act}}\right|^2}{\left(\omega_{\mathrm{eff}}^2-\omega^2\right)^2+\left(\omega \Gamma_{\mathrm{eff}}\right)^2},    
\end{align}
where the amplitude of the actuation laser is assumed to be sufficiently large to make the thermal noise negligible.

\section{On the dip structure appearing in the cavity length dependence of LIOFs}\label{Appendix:G}
\begin{figure}[htbp] % CCCCCCCCCCCCCCCCCCCCCC
        \centering
          \includegraphics[width=180mm]{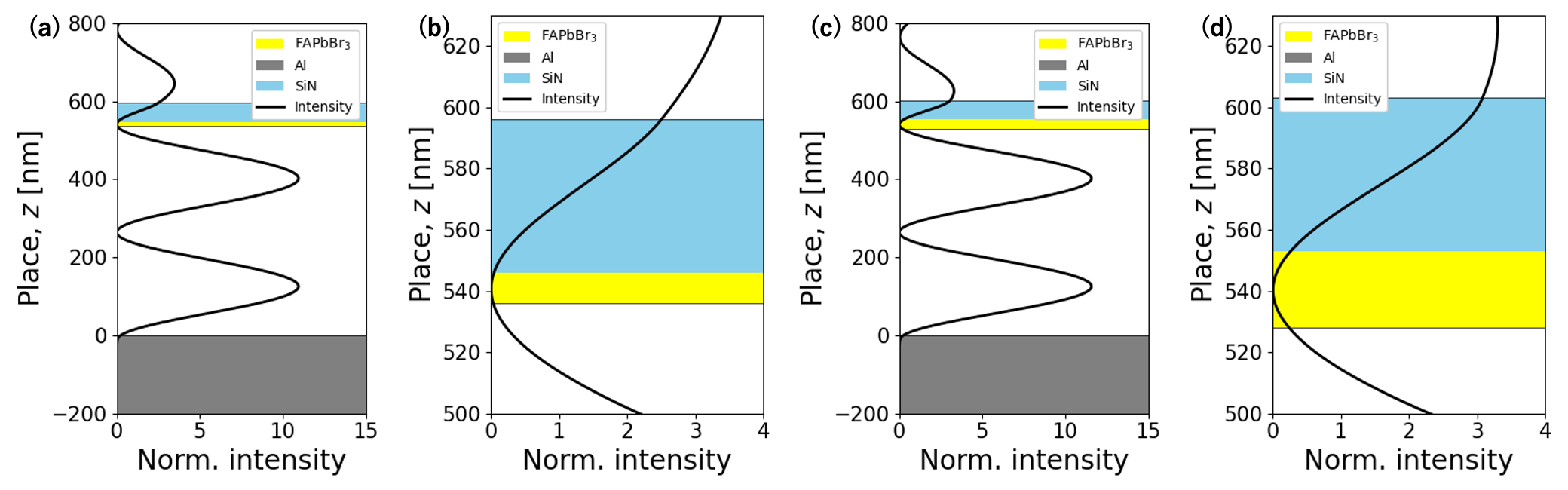}
    \caption{
    \bla{Calculated spatial distribution of excitation light intensity with luminescent film thicknesses of $d=10$ nm in panel (a,b) and $d=25$ nm in panel (c,d). (a,b) are the results for a cavity length of $L=536$ nm and (c,d) are the conditions for $L=528$ nm. Panels (b) and (d) are magnified views of (a) and (c), respectively, to show the distribution of electric field intensity within the emitter film more clearly. The blue, yellow, and gray regions represent SiN, $\text{FAPbBr}_{3}$, and Al, respectively.}
        \label{EF_dist}
        }
\end{figure}

\bla{We calculated the spatial distribution of the excitation light intensity under conditions where the LIOF dip occurs, specifically at the cavity length of $L=536, 528$ nm in the case of film thicknesses $d=10, 25$ nm, as shown in Fig. \ref{EF_dist}.
Figure 2(d) in the main text displays the induced force as a function of $L$ for different film widths. The dip structures of LIOFs appear when the luminescent film is thin. This dip occurs because the emitter is positioned at the nodes of the standing wave of the excitation light, as depicted in Fig. \ref{EF_dist}. Consequently, the emitter is not sufficiently excited, leading to a lack of light emission and a localized weakening of the LIOF. }
%%%%%%% Figure ? %%%%%%%

%\input{Appendix_A.tex}
%\input{Appendix_B.tex}
%\input{Appendix_C.tex}
%\input{Appendix_D.tex}
%\input{Appendix_E.tex}

%%%%%%Bibliography%%%%%%
\bibliography{bibtex_2025.bib}

%%%%%%Appendix%%%%%%

\end{document}